\documentclass[11pt]{article}
\usepackage{graphicx}
\usepackage{epsfig}
\usepackage{subfigure}

\newcommand{\BABARPubYear}    {00}
\newcommand{\BABARConfNumber}  {08}
\newcommand{\SLACPubNumber} {8530}

\newcommand{\deltae}{\ensuremath{\Delta E}}

\def\D       {\ensuremath{D}}

\def\Dm      {\ensuremath{D^-}}
\def\Dstar   {\ensuremath{D^{*}}}

\def\Dstarm  {\ensuremath{D^{*-}}}

\def\DDstarm {\ensuremath{D^{(*)-}}}
\def\rhop    {\ensuremath{\rho^+}}
\def\rhom    {\ensuremath{\rho^-}}
\def\aonep   {\ensuremath{a_1^+}}

\usepackage{relsize}
\def\babar{\mbox{\slshape B\kern-0.1em{\smaller A}\kern-0.1em
    B\kern-0.1em{\smaller A\kern-0.2em R}}}

\def\epem       {\ensuremath{e^+e^-}}

\def\piz   {\ensuremath{\pi^0}}

\def\pip   {\ensuremath{\pi^+}}
\def\pim   {\ensuremath{\pi^-}}

\def\Kbar  {\kern 0.2em\overline{\kern -0.2em K}{}}

\def\Kp    {\ensuremath{K^+}}

\def\KS    {\ensuremath{K^0_{\scriptscriptstyle S}}} 
 
\def\Kstarz  {\ensuremath{K^{*0}}}

\def\Kzb   {\ensuremath{\Kbar^0}}
\def\KzKzb {\ensuremath{K^0 \kern -0.16em \Kzb}}
\def\Dz    {\ensuremath{D^0}}
\def\Dbar  {\kern 0.2em\overline{\kern -0.2em D}{}}

\def\Dzb   {\ensuremath{\Dbar^0}}
\def\DzDzb {\ensuremath{D^0 {\kern -0.16em \Dzb}}}
\def\Dstar   {\ensuremath{D^*}}

\def\Bz    {\ensuremath{B^0}}
\def\B     {\ensuremath{B}}
\def\Bbar  {\kern 0.18em\overline{\kern -0.18em B}{}}

\def\Bzb   {\ensuremath{\Bbar^0}}

\def\BB    {\ensuremath{B\Bbar}} 
\def\BzBzb {\ensuremath{B^0 {\kern -0.16em \Bzb}}}

\def\jpsi  {\ensuremath{{J\mskip -3mu/\mskip -2mu\psi\mskip 2mu}}} 
\mathchardef\Upsilon="7107
\def\Y#1S{\ensuremath{\Upsilon{(#1S)}}}

\def\FourS {\Y4S}

\mathchardef\Deltares="7101
\mathchardef\Xi="7104
\mathchardef\Lambda="7103
\mathchardef\Sigma="7106
\mathchardef\Omega="710A
\def\Deltabar   {\kern 0.25em\overline{\kern -0.25em \Deltares}{}}
\def\Lbar {\kern 0.2em\overline{\kern -0.2em\Lambda\kern 0.05em}\kern-0.05em{}}
\def\Sigbar{\kern 0.2em\overline{\kern -0.2em \Sigma}{}}
\def\Xibar{\kern 0.2em\overline{\kern -0.2em \Xi}{}}
\def\Obar{\kern 0.2em\overline{\kern -0.2em \Omega}{}}
\def\Nbar{\kern 0.2em\overline{\kern -0.2em N}{}}
\def\Xbar{\kern 0.2em\overline{\kern -0.2em X}{}}

\def\mes        {\mbox{$m_{\rm ES}$}}

\def\ev   {\ensuremath{\rm \,e\kern -0.08em V}}
\def\kev  {\ensuremath{\rm \,ke\kern -0.08em V}} 
\def\mev  {\ensuremath{\rm \,Me\kern -0.08em V}} 
\def\gev  {\ensuremath{\rm \,Ge\kern -0.08em V}} 
\def\gevc {\ensuremath{{\rm \,Ge\kern -0.08em V\!/}c}} 
\def\tev  {\ensuremath{\rm \,Te\kern -0.08em V}}
\def\mevc {\ensuremath{{\rm \,Me\kern -0.08em V\!/}c}} 
\def\gevcc{\ensuremath{{\rm \,Ge\kern -0.08em V\!/}c^2}} 
\def\mevcc{\ensuremath{{\rm \,Me\kern -0.08em V\!/}c^2}}

\def\cm   {\ensuremath{\rm \,cm}}

\def\mm   {\ensuremath{\rm \,mm}}

\def\mum  {\ensuremath{\,\mu\rm m}} 

\def\invfb   {\ensuremath{\mbox{\,fb}^{-1}}}
\def\mus  {\ensuremath{\rm \,\mus}}

\def\ps   {\ensuremath{\rm \,ps}}

\def\mus        {\ensuremath{\,\mu{\rm s}}}    
\def\ps         {\ensuremath{{\rm \,ps}}}   
\def\mrad{\ensuremath{\rm \,mr}}                

\def\gsim{{~\raise.15em\hbox{$>$}\kern-.85em
          \lower.35em\hbox{$\sim$}~}}
\def\lsim{{~\raise.15em\hbox{$<$}\kern-.85em
          \lower.35em\hbox{$\sim$}~}}

\def\CP                 {\ensuremath{C\!P}}

\def\to                 {\ensuremath{\rightarrow}}

\def\pep2{PEP-II}

\newcommand{\dedx}{\ensuremath{\mathrm{d}\hspace{-0.1em}E/\mathrm{d}x}}

\def\mistag{\ensuremath{w}}

\def\deltaz{\ensuremath{{\rm \Delta}z}}
\def\deltat{\ensuremath{{\rm \Delta}t}}
\def\deltamd{\ensuremath{{\rm \Delta}m_d}}

\newcommand{\eqref}[1]{Eq.~(\ref{eq:#1})}

\newcommand{\epjc}      [1]  {{Eur.\ Phys.\ Jour.\ C~{\bf #1}}}

\newcommand{\pl}        [1]  {{Phys.\ Lett.\ {\bf #1}}}      

\newcommand{\prl}       [1]  {{Phys.\ Rev.\ Lett.\ {\bf #1}}} 

\newcommand{\zp}        [1]  {{Z.\ Phys.\ {\bf #1}}}

\def\jetset74   {\mbox{\tt Jetset \hspace{-0.5em}7.\hspace{-0.2em}4}}

\providecommand{\xpm}{\mbox{$\pm$}}

\providecommand{\btodstarlnu}{\mbox{$B\to D^{*}l\nu$}}

\providecommand{\B}{\mbox{$B$}}
\providecommand{\dt}{\mbox{$\Delta t$}}

\long\def\inst#1{\par\nobreak\kern 4pt\nobreak
    {\it #1}\par\vskip 10pt plus 3pt minus 3pt}

\begin{document}
{\pagestyle{empty}

\begin{flushright}
\babar-CONF-\BABARPubYear/\BABARConfNumber \\
SLAC-PUB-\SLACPubNumber
\end{flushright}




\par\vskip 3cm

\begin{center}
\Large \bf A measurement of the \boldmath \BzBzb\ oscillation frequency 
and determination of flavor-tagging efficiency using\\
semileptonic and hadronic \Bz\ decays 
\end{center}
\bigskip

\begin{center}
\large The \babar\ Collaboration\\
\mbox{ }\\
July 25, 2000
\end{center}
\bigskip \bigskip

\begin{center}
\large \bf Abstract
\end{center}

\noindent
\BzBzb\ flavor oscillations are studied in
\epem\ annihilation data collected with the
\babar\ detector at center-of-mass energies near the \FourS\ resonance.
One $B$ is reconstructed in a hadronic or semileptonic decay mode, and
the flavor of the other $B$ in the event is determined with a tagging 
algorithm that exploits the relation between the flavor of the 
heavy quark and the charges of its decay products. 
Tagging performance is characterized by an efficiency $\epsilon_i$ and
a probability for mis-identification, $\mistag_i$, for each tagging category.
We report a determination of the wrong-tag
probabilities, $\mistag_i$, and 
a preliminary result for the 
time-dependent \BzBzb\ oscillation frequency,
$\deltamd = 0.512 \xpm\  0.017 \xpm\ 0.022$~$\hbar$\ps$^{-1}$.

\vfill
\centerline{Submitted to the XXX$^{th}$ International Conference on High Energy Physics, Osaka, Japan.}
\newpage
}

\newcommand{\secname}{}

\begin{center}
\small

The \babar\ Collaboration
\bigskip

B.~Aubert,
A.~Boucham,
D.~Boutigny,
I.~De Bonis,
J.~Favier,
J.-M.~Gaillard,
F.~Galeazzi,
A.~Jeremie,
Y.~Karyotakis,
J.~P.~Lees,
P.~Robbe,
V.~Tisserand,
K.~Zachariadou
\inst{Lab de Phys.\ des Particules, F-74941 Annecy-le-Vieux, CEDEX, France}
% ...

\end{center}\newpage

\setcounter{footnote}{0}

\renewcommand{\secname}{Introduction}
\section{Introduction}
\label{sec:Introduction}  

The phenomenon of particle--anti-particle mixing 
in the neutral $B$ meson system was first observed almost fifteen years
ago~\cite{UA1},~\cite{ARGUS}.  
The oscillation frequency in \BzBzb\ mixing\footnote[1]{The symbol $B^0$ refers 
to the $B_d$ meson; charge conjugate
modes are implied throughout this paper.}
has been extensively studied with both time-integrated  and 
time-dependent techniques~\cite{PDG}.  
In this paper we present a preliminary measurement of time-dependent mixing performed 
at the PEP-II asymmetric $e^+e^-$ collider, where resonant production of
the \FourS\  provides a 
copious source of \BzBzb\ pairs moving along the beam axis ($z$ direction)   
with a Lorentz boost of $\beta\gamma = 0.56$.
The typical separation between the two $B$ decay vertices is
$\deltaz = \beta\gamma c \tau_B = 260$\mum, where $\tau_B =
1.548\pm 0.032$\ps\ is the \Bz\ lifetime~\cite{PDG}.
The \BzBzb\ mixing probability
is a function of \deltamd, the difference between 
the mass eigenstates $B^0_H$ and $B^0_L$,  and the 
time between the $B$ decays, $\deltat = \deltaz/\beta\gamma c$:

\begin{equation}
Prob(\Bz \rightarrow \Bzb) \propto {\rm{e}}^{-|\dt|/\tau_B}(1 - \cos \deltamd \deltat).
\label{eq:mix} 
\end{equation}

In the Standard Model,
\BzBzb\ mixing occurs through second-order weak diagrams
involving the exchange of up-type quarks, 
with the top quark contributing
the dominant amplitude.  A measurement of \deltamd\ is therefore
sensitive to the value of the Cabbibo-Kobayashi-Maskawa matrix~\cite{KM} element
$V_{td}$.  At present the sensitivity to $V_{td}$ is not limited by
experimental precision on \deltamd, but by other
uncertainties in the calculation, in particular the quantity
$f_B^2 B_B$, where $f_B$ is the \Bz\ decay constant, and 
$B_B$ is the so-called bag factor, 
representing  the strong interaction matrix elements.

In this analysis, we study the time-dependent probability to observe
\BzBzb, \Bz\Bz\ and \Bzb\Bzb\ pairs produced in
\FourS\ decay.  We reconstruct
one $B$ in a flavor eigenstate, and use the remaining particles from
the decay of the other $B$ to identify, or ``tag'', its flavor.
The charges of identified leptons and kaons are the primary indicators of the flavor of
the tagging $B$, but other particles also carry flavor information that
can be identified with a  neural network algorithm.
The tagging algorithm used in this analysis is identical to that
employed by \babar\ in \CP\ violation studies, 
in which one $B$ is fully reconstructed in a
\CP\ eigenstate~\cite{BabarPub0001}.  

Considering the \BzBzb\ system as a whole, one can classify the tagged events 
as {\em mixed} or {\em unmixed} depending on whether the reconstructed $B$,
referred to as $B_{rec}$, has the same or the opposite flavor as the 
tagging $B$,
referred to as $B_{tag}$. 
If the flavor tagging  were perfect, 
the asymmetry 
\begin{equation}
a(\deltat) = \frac{N_{\rm unmix}-N_{\rm mix}}{N_{\rm unmix}+N_{\rm mix}} 
\label{eq:asym}
\end{equation}
plotted as a function of \deltat\ would describe a cosine function with unit amplitude.
However, the tagging algorithm incorrectly identifies the tag with a 
probability $\mistag$.  This mistag rate reduces the
amplitude of the oscillation by a ``dilution
factor'' ${\cal D} = (1 -2\mistag)$.  When more than one type of flavor tag is
employed, each will have its own mistag
rate, $\mistag_i$.
A simultaneous fit to the mixing frequency
and its amplitude allows the determination of both \deltamd\
and the mistag rates, $\mistag_i$.  

Neglecting any background contributions, 
the probability density functions (PDF's) 
for the mixed $(-)$ and unmixed $(+)$ events can be expressed as
the convolution of the oscillatory component $h_{\pm}$, with a time resolution
function ${\cal {R}}(\dt|\hat {a})$:
\begin{equation}
\label{eq:pdf}
{\cal H}_\pm(\, \dt; \Gamma, \, {\rm \Delta} m_d, \, {\cal {D}}, \, \hat {a} \, )  = h_\pm( \, \dt; \Gamma, \, {\rm \Delta} m_d, \, {\cal {D}} ) 
\otimes 
{\cal {R}}( \dt ; \hat {a} ) ,
\end{equation}
where $\hat a$ are the parameters of the resolution function and
\begin{equation}
        h_\pm(\, \dt; \Gamma, \, {\rm \Delta} m_d, \, {\cal {D}}\, )  = {\frac{1}{4}}\, \Gamma \, {\rm e}^{ - \Gamma \left| \dt \right| }
\, \left[  \, 1 \, \pm \, {\cal {D}}  \cos{ {\rm \Delta} m_d \, \dt } \,  \right] .
\end{equation}
The log-likelihood function is then constructed from the sum of ${\cal {H}}_{\pm}$ over all mixed
and unmixed events, and 
over the different tag types,
$i$, each with its own characteristic dilution factor ${\cal {D}}_i$:
\begin{equation}
\label{eq:Likelihood_eq}
 {\ln {\cal {L} } } = \sum_{ i}^{\rm tagging} \left[ \, \sum_{{\rm unmixed}}{  \ln{ {\cal H}_+( 
\, \dt; \Gamma, \, {\rm \Delta} m_d, 
\, {\cal {D}}_i, \,   \hat {a}_i  \,) } } + \sum_{ {\rm mixed} }{ \ln{ {\cal H}_-( 
\, \dt ; \Gamma, \, {\rm \Delta} m_d, 
\, {\cal {D}}_i, \,  \hat {a}_i  \, ) } } \, \right].
\end{equation}

The log-likelihood is maximized to extract the dilutions
${\cal D}_i$ and, simultaneously, the mixing parameter \deltamd.
The correlation between these parameters is small, because
the rate of mixed events at low values of \deltat, where
the \BzBzb\ mixing probability is small, is principally 
governed by the mistag rate. Conversely, the sensitivity 
to \deltamd\ increases at larger values of \deltat;
when \deltat\ is approximately twice the $B$ lifetime, half of
the neutral $B$s will have oscillated.

Alternatively, the mistag rate can be extracted in a 
time-independent analysis.  Neglecting 
possible background contributions and assuming the $B_{rec}$ flavor is 
correctly identified,
one can express the observed time-integrated fraction of 
mixed events $\chi_{obs}$ as a function of the \BzBzb\ mixing probability  $\chi_d$~:
\begin{equation}
\label{eq:TagMix:Integrated}
\chi_{obs} = \chi_d  + (1 - 2 \chi_d )\, \mistag,
\end{equation}
where 
$\chi_d = \frac{1}{2} \, x_d^2 / ( 1+ x_d^2 )$ 
and $ x_d = {\rm \Delta} m_d /\Gamma $.
The current world average for $\chi_d$ is $0.174 \pm 0.009$~\cite{PDG}. 
Because decay time information is available in \babar, we can improve
the statistical precision on $\mistag$ by selecting only events that 
fall into an optimized time interval, $|\dt| < t'$, where $t'$ is
the value of $|\deltat|$ above which the integrated number of
mixed events equals the integrated number of unmixed events.
Through the use of an optimized $\dt$ interval
this method achieves nearly the same statistical precision 
on $\mistag$ that is obtained in a full
time-dependent likelihood fit.

\renewcommand{\secname}{Detector}
\section{The \babar\ detector and data set}
\label{sec:Detector}  

The data used in this analysis were collected with the \babar\ 
detector~\cite{BabarPub0017}
at the \pep2\ storage ring in the period  January--June, 2000.
The total integrated luminosity of the data set is 8.9\invfb\
collected near the \FourS\ resonance and
0.8\invfb\ collected 40\mev\ below the \FourS\ resonance
(off-resonance data). The corresponding number
of produced \BB\ pairs is estimated to be about $(10.1\pm 0.4) \times 10^6$.

The \babar\ detector is described in more detail elsewhere~\cite{BabarPub0017}.
For this analysis, the most important detector capabilities include
charged-particle tracking, vertexing and particle identification.
Charged particles
are detected and their momenta  measured by a combination of 
a central drift chamber (DCH) filled with a helium-based gas  and  a
five-layer, double-sided silicon vertex tracker (SVT), immersed
in a 1.5~T axial field produced by a superconducting magnet.  The 
charged-particle momentum resolution is  given by 
$\left( \delta p_T / p_T \right)^2 =  (0.0015\, p_T)^2 + (0.005)^2$, where 
$p_T$ is in \gevc.  The SVT, 
with typical 10\mum\ single-hit resolution, 
provides vertex information in both  
the transverse plane and in $z$.  The $B$ meson decay 
vertex resolution is typically 50\mum\ in $z$  for 
a fully reconstructed \B\ meson and about  
100 to 150\mum\ for the companion (unreconstructed)  \B\ meson in the event.
Beyond the outer radius of the DCH is a detector of internally reflected 
Cherenkov radiation (DIRC) which is used primarily for charged hadron
identification. The device consists of quartz bars in
which Cherenkov light is produced as relativistic charged particles traverse
the material. The light is internally reflected, and the Cherenkov rings
are measured with an array of photomultiplier tubes mounted on the rear of
the detector. A CsI(Tl) crystal electromagnetic calorimeter (EMC) is used to 
detect photons, and neutral hadrons, and also for electron
identification. The EMC is surrounded by a superconducting 
coil which provides a 1.5 T magnetic field. The Instrumented Flux Return
(IFR) consists of multiple layers of resistive plate chambers 
interleaved with the flux return iron and is
used in the identification of muons and neutral hadrons.


\renewcommand{\secname}{PID}
\section{Particle identification}
\label{sec:PID}  


Identification of electrons, muons and kaons is an essential
ingredient of both $B$ reconstruction and flavor tagging. As noted above, the \babar\
detector has several systems that contribute to particle
identification, including the measured \dedx\ in the SVT and the drift chamber,
Cherenkov angle determination in the DIRC, electromagnetic energy
measurement in the CsI calorimeter and detection of penetrating
particles in the IFR.  In this section we describe in detail the
selections that are employed to identify particles by species, both for
the purposes of flavor tagging and for $B$ reconstruction, where the
latter typically employs looser selection criteria.  We also include
the preliminary average values for efficiency and pion
misidentification probabilities, as determined from data.

\subsection{Electron identification}
For tagging purposes, an electron candidate must be matched to an
electromagnetic cluster in the CsI calorimeter consisting of at least three
crystals, and the ratio of the cluster energy to the track momentum
$E/p$ must be between 0.88 and 1.3. Cuts~\cite{BabarPub0017} based on electromagnetic 
shower shape are also used in identifying electron candidates.
In addition the electron candidate track is  required to have a specific ionization (\dedx)
measurement in the drift chamber  consistent
with that of an electron, and, if measured, the Cherenkov angle in the
DIRC is required to be consistent with that of an ultra-relativistic particle.
The electron efficiency and pion misidentification probabilities for this
selection are about 92\% and 0.3\%, respectively.

A somewhat looser electron selection is used in lepton identification for
$B^0\rightarrow D^{*-}e^+\nu$ events. In this case, the \dedx\
and $E/p$ selections are loosened, while the
requirements on electromagnetic shower shape and Cherenkov angle are removed. This
looser selection has efficiency for electrons of approximately 97\%,
while the pion misidentification probability is about 3\%. 

\subsection{Muon identification}
Muon identification relies mainly on the number of measured
interaction lengths $\lambda$ traversed by the track in the IFR iron.
At least $2.2\lambda$ are required, and at higher momenta we require
more
than $\lambda_{exp}-1$, where $\lambda_{exp}$ is the number of
expected interaction lengths as a function of momentum.  
To reject hadronic showers, we make requirements
on the number of hit IFR strips that are parametrized as a function of the
penetration length and the distance of the hits from the extrapolated track.  
In the forward region, which suffers from machine
background, extra hit-continuity criteria are applied.  In addition,
if the particle is in an angular region covered by the EMC, the
muon candidate must have an energy deposit in the calorimeter larger
than 50\mev\ and smaller than 400\mev. 

These same selection criteria are
used for both tagging purposes and muon identification in the decay $B^0\rightarrow
D^{*-}\mu^+\nu$. The average efficiency and pion misidentification
probability are about 75\% and 2.5\%, respectively.

\subsection{Kaon identification}
Kaon identification employed for flavor tagging requires the ratio of
the combined kaon likelihood to the combined pion likelihood be
larger than 15.  The combined likelihoods are obtained by multiplying
the individual
likelihoods from the SVT, DCH and DIRC subsystem information.  In the SVT and DCH
tracking detectors, the likelihoods
are based on the \dedx\ truncated mean measurement
compared to the expected value for the $K$ and $\pi$ hypotheses,
assuming a Gaussian distribution.  The \dedx\
resolution is estimated on a track by track basis given the
direction and momentum of the track and the number of energy
deposition samples.
In the DIRC, the likelihood is computed by combining the likelihoods computed from
the measured Cherenkov angle, in comparison with the expected Cherenkov angle
for a given mass hypothesis, and the Poisson probability for the
observed number of Cherenkov photons, given the number of expected photons for
the same hypothesis.  DIRC information is not required below 0.7\gevc\
where the DCH \dedx\ alone provides good $K / \pi$
discrimination.  The efficiency of this selection for kaons is about
85\% and the pion misidentification probability is about 5\%.

In the reconstruction of \Bz\ decays, some channels require a 
very loose kaon selection to reduce
backgrounds to acceptable levels.  The selection criteria are similar those
described above, except that the kaon hypothesis is assumed for those subsystems that
have no particle identification
information.
This looser selection results in a higher kaon efficiency of about 95\%
and an acceptable pion misidentification probability of
20\%. In addition, a loose pion selection is employed in
reconstructing some $B$ decay modes. This requirement consists of
rejecting pion candidates if they satisfy the tighter kaon or lepton
criteria used for flavor tagging described above.

\renewcommand{\secname}{Vertex}
\section{Time resolution function}
\label{sec:\secname}
\par
The time difference, $\Delta t = t_{rec} - t_{tag}$, between $B$ decays is determined from the
 measured separation $\Delta z$ 
between the reconstructed $B$ and flavor-tagging decay 
vertices along the $z$ axis using the known boost, $\Delta t = \Delta z/\beta\gamma c$.
The $\Delta t$ resolution 
is dominated by the $z$ position resolution for the flavor-tagging $B$ meson vertex.
The $B_{tag}$ production point and three-momentum, with its associated 
error matrix, are derived from the fully reconstructed $B_{rec}$ candidate 
three momentum, decay vertex and error matrix, and from the knowledge of 
the average position of the interaction point and the \FourS\ average boost.
These $B_{tag}$ parameters are used as input to a
geometrical fit to a single vertex, 
including all other tracks in the event except those used to reconstruct 
$B_{rec}$. In order to reduce 
bias due to long-lived particles, all possible $V^0$ 
candidates that can be reconstructed are used as input to the fit in place
of their daughters.  Tracks whose contributions to the $\chi^2$
are greater than 6 are removed from the fit, 
in an iterative procedure that continues until all remaining tracks satisfy this requirement 
or all tracks are removed. 

The time resolution function 
can be approximated by a sum of three Gaussian distributions with different means and widths
\begin{eqnarray}
{\cal {R}}( \, \deltat ; \, \hat {a} \,  ) &=&  \sum_{i=1}^{3} { \, \frac{f_i}{\sigma_i\sqrt{2\pi}} \, {\rm exp} \left(  - ( \deltat-\delta_i )^2/2{\sigma_i }^2   \right) } \, \, .
\end{eqnarray}
\par
Fitting the vertex resolution function with simulated events shows 
that most of the events ($f_1 = 1-f_2-f_3 \approx 70\%$) are found in a core Gaussian component, 
which has a width $\sigma_1 \approx 0.6 \ps$, while the remaining events
reside in the tail Gaussian, which has a width $\sigma_2 \approx 1.8\ps$. 
Tracks from forward-going charm decays included in the reconstruction of the $B_{tag}$ vertex introduce a small 
bias, $\delta_1 \approx -0.2 \ps$, for the core Gaussian towards negative values of \deltat.  

A small fraction of events have large values of $|\deltat |$, due to incorrectly reconstructed
vertices.  This is  accounted for in the 
parameterization of the time resolution function by the third Gaussian component,
centered at zero with broad fixed width of $8\ps$, making it almost constant over the time interval of the fit.
The fraction of events populating this component of the resolution function, $f_{w}\equiv f_3$
is approximately 2 \%.
 
In the final likelihood fits, we describe the 
$\Delta t$ resolution by introducing two scale factors 
${\cal S}_1$ and ${\cal S}_2$ that are applied to the
event-by-event resolution,
$\sigma_{\deltat}$, calculated the error on \deltaz\ provided 
by the vertex fit.
We take the width of the core and the tail Gaussian components for 
each event to be $\sigma_1={\cal S}_1 \times \sigma_{\deltat}$ and 
$\sigma_2={\cal S}_2 \times \sigma_{\deltat}$, respectively.
The scale factor ${\cal S}_1$ and the bias $\delta_1$ of the 
core Gaussian are free parameters in the fit.  
The scale factor ${\cal S}_2$ and the fraction of events in the core
Gaussian $f_1$ are fixed to the values estimated 
from Monte Carlo simulation by a fit to the pull distribution (${\cal S}_2=2.1$
and $f_1=0.75$). The bias of the tail Gaussian, $\delta_2$, 
is fixed at $0 \ps$. The three free parameters in the likelihood fit are
\begin{eqnarray} 
\hat {a} &=& \{ \, {\cal S}_1, \,  \delta_1, \,  f_{w}  \} \, \, .
\end{eqnarray}

We observe no significant differences in simulated events between
resolution function parameters obtained from samples involving
different decay modes for $B_{rec}$. This is  
expected, because $\Delta t$ resolution is dominated by the 
precision on the $B_{tag}$, rather than the $B_{rec}$ vertex. 
Likewise, the differences in the resolution function 
parameters for the different tagging categories are also small.  
Therefore, 
we use a single set of resolution function 
parameters $\hat {a}$ for all decay modes. 
Table~\ref{tab:Resolution} shows the values for the vertex parameters
obtained in data from a fit to the hadronic \Bz\ sample, described below.

\begin{table}[!htb]
\caption{
Parameters of the resolution function determined from the sample
of events with hadronic fully reconstructed $B$ candidates.
} 
\vspace{0.3cm}
\begin{center}
\begin{tabular}{|cc|cl|} \hline
   \multicolumn{2}{|c|}{parameter} & \multicolumn{2}{c|} {value}    \\ \hline \hline
 $\delta_1$  & (ps)    & $-0.20\pm0.07$  & from fit     \\
 ${\cal S}_1$&   & $1.33\pm0.13$       & from fit     \\
 $f_{w}$       & (\%)  & $1.6\pm0.6$     & from fit     \\
 $f_1$       & (\%)  & $75$              & fixed        \\
 $\delta_2$  & (ps)  & $0$             & fixed        \\
 ${\cal S}_2$ &  & $2.1$               & fixed        \\
\hline
\end{tabular}
\end{center}
\label{tab:Resolution}
\end{table}

\renewcommand{\secname}{Tagging}
\section{Flavor tagging}
\label{sec:Tagging}

After the daughter tracks of the reconstructed $B$ are removed,
the remaining tracks are analyzed to determine the flavor of the
$B_{tag}$, and this ensemble is assigned a tag flavor, either $\Bz$ or $\Bzb$.  

To illustrate the tagging discriminating power of each tagging category,   
we use as a figure of merit the effective tagging efficiency  
$Q_i = \epsilon_i \times \left( 1 - 2\mistag_i \right)^2$, where $\epsilon_i$ 
is the fraction of events associated to the tagging category $i$ and 
$\mistag_i$ is the mistag fraction, the probability of incorrectly assigning 
the opposite tag to an event of this category. 

We use four different types of flavor tag, or tagging categories,  
in this analysis.
Two of these tagging categories
rely upon the presence of a prompt lepton or 
one or several charged kaons in the event.  
The remaining two categories, called neural network categories, are based upon
the output values of a neural network algorithm applied to all the events 
that have not already been assigned to one of the {\tt Lepton} or {\tt Kaon}  
tagging categories.
   
\subsection{Lepton and kaon tags}
\label{sec:\secname:NOT}
\par
The  {\tt Lepton} and {\tt Kaon} tagging categories  
use the correlation
between the charge of a primary lepton from a semileptonic decay or
the charge of a kaon, and the flavor of the decaying $b$ quark.  
For the {\tt Lepton} category we use both electrons and muons.
A minimum center-of-mass momentum requirement 
on the lepton is applied to reduce
the contamination from softer opposite-sign leptons coming from 
charm semileptonic decays.  There are no such discriminating 
kinematic quantities to reduce the contamination of opposite-sign kaons,
so  the optimization relies principally  on 
the balance between the kaon identification efficiency and the purity of 
the kaon sample.  

A lepton tag is
defined by taking the charge of the fastest identified electron or muon 
with a center of mass momentum greater than 1.1\gevc. 
A kaon tag is defined by taking the sum of the charges of all identified 
kaons.  If both an electron and a muon are identified, the electron
tag takes precedence.  
If the event has a lepton tag and there is no conflicting kaon tag,
the event is assigned to the {\tt Lepton} category.
If the event has no lepton tags but has a non-zero kaon
tag, the event is assigned to the {\tt Kaon} category.
If the event has both lepton and kaon tags but they conflict, the
event is not assigned to either the {\tt Lepton} or the {\tt Kaon} category.
  
\subsection{Neural network tags}
\label{sec:\secname:NetTagger}
\par 
The use of a 
second tagging algorithm
is motivated by the 
potential flavor-tagging power carried by non-identified 
leptons and kaons, softer leptons from charm semileptonic decays, 
soft pions from $D^*$ decays, and more generally, by the momentum 
spectrum of charged particles from $B$ meson decays.  
The best way to exploit the information contained in a set of correlated
quantities is to use multivariate methods, such as neural networks, 
rather than to apply selection cuts.

We design five different neural networks, called subnets, each with 
a specific goal.  Four  of the  subnets are track-based: the 
{\tt L} and {\tt LS} subnets are sensitive to the presence
of primary and cascade leptons respectively, the {\tt K} subnet 
to that of charged kaons and the  {\tt SoftPi} subnet to that 
of soft pions from $D^*$ decays.  In addition the  {\tt Q} subnet 
exploits the charge of high-momentum particles in the event.

The {\tt L} and {\tt LS} subnets share a set of
discriminating kinematic variables in addition to the boolean outputs of 
the standard lepton identification algorithms.
The variables are designed to more fully characterize 
the kinematic features of a true primary lepton from a semileptonic $B$ decay. 
Taking all charged tracks in the event, excluding that under consideration as the lepton, 
together with all neutral clusters we form the recoiling system $X$.
Assuming that $B_{tag}$ is produced at rest in the center-of-mass frame, 
and that 
the excluded track is a primary lepton from the $B_{tag}$ semileptonic decay,
one can calculate  
the four-momentum of the neutrino $p_{\nu}$ and the four-momentum 
$p_W$ of the virtual $W$ boson in the semileptonic decay.
The  variables available
are the center-of-mass momentum of the lepton candidate track, the mass and energy of the recoil system $X$,
the cosine of the angle between the excluded track momentum and 
the direction of the 
neutrino, 
the cosine of the angle between the $W$ momentum direction and the direction of 
the closest charged or neutral candidate with energy larger than 50\mev, 
and the 
total energy flow in the hemisphere around the $W$ momentum direction.  

The {\tt K} subnet uses the particle momentum in the laboratory frame, together with the 
three relative likelihoods $L_K/( L_{\pi} + L_K)$ for the SVT, the DCH and the DIRC.

The {\tt SoftPi} subnet uses the track 
with the lowest momentum  from  the $B_{tag}$ decay 
and the angle it makes with respect to the thrust axis, 
calculated using all other charged tracks.

The variables of the {\tt Q} subnet
are the momentum of the highest momentum track in the  $B_{tag}$  system 
and the sum of momentum-weighted charges
normalized by the momentum sum of all tracks with impact parameter less than
1\mm\ and all neutral clusters with energy greater than 50\mev. 

The variables used as input to the  
neural network tagger are the highest values of 
the {\tt L}, {\tt LS} and {\tt SoftPi} subnet outputs each multiplied by the charge
of the corresponding tagging track, 
the highest and the second-highest values of the {\tt K} subnet
output again multiplied by the charge of the corresponding tagging tracks,
and the output of the {\tt Q} subnet.

The output from the full neural network tagger, $x_{NT}$, can be mapped onto the interval $\left[ -1, 1 \right]$.  
The assigned flavor tag is \Bz\ if $x_{NT}$ is negative, and \Bzb\ otherwise.
Events with $\left| x_{NT} \right| > 0.5$ are assigned to  the {\tt NT1} tagging category 
and events with  $0.2 < \left| x_{NT} \right| < 0.5$ to the {\tt NT2} tagging category.
Events with  $\left| x_{NT} \right| < 0.2$  have very little tagging 
power and are rejected.

\renewcommand{\secname}{Sample}
\section{Event selection and sample composition}  
\label{sec:Sample}  

\Bz\ mesons are reconstructed in the  hadronic decay modes 
$\Bz \to D^{(*)-} \pi^+$, $D^{(*)-} \rho^+$, $D^{(*)-} a_1^+$,
\jpsi \Kstarz\ and the semileptonic decay mode $\Bz \to
D^{*-}\ell^+\nu$.
All final state particles, with the exception of the neutrino in
the semileptonic decay, are reconstructed.  

Charged tracks measured by the drift chamber and/or SVT
are required to originate within 1.5\cm\ in $xy$ and 10\cm\ in $z$
of the nominal beamspot. Electromagnetic
clusters in the calorimeter, that are unassociated with charged tracks and
used to reconstruct $\pi^0$ candidates, are required to have an energy greater than 30\mev\ and
a shower shape consistent with a photon interaction.

Neutral pion candidates are formed from pairs of electromagnetic
clusters assumed to be photons. The invariant mass of the
photon pair is required to be within $\pm 20$\mevcc\ (2.5 $\sigma$ ) of the nominal \piz\
mass, with a minimum summed energy of 200\mev. Selected
candidates are kinematically fitted with a \piz\ mass constraint.

\KS\ candidates are reconstructed in the \pip\pim\ mode, with an
invariant mass, computed at the vertex of the two tracks, between 462
and 534\mevcc. The $\chi^2$ of the topological vertex fit is required to have a 
probability greater than 0.1\%. The opening angle between the
flight direction and the momentum vector 
for the \KS\ candidate is required to be smaller than 200\mrad. 
Finally, the transverse flight distance
from the primary vertex in the event, $r_{xy}$, is required to be greater than 2\mm.

In order to reject ``jet-like'' events from 
$ e^+e^- \rightarrow q\overline q$ (continuum) background,
we require that the normalized second Fox-Wolfram moment~\cite{fox} ($R_2=H_2/H_0$), 
calculated with both charged tracks and neutral clusters, be less 
than 0.5 (0.45) in hadronic (semileptonic) decay modes. 

\subsection{Hadronic \boldmath \Bz\ decays}
\label{subsec:Sample_hadronic}

\Bz\ mesons are reconstructed in the hadronic modes $\Bz\to\DDstarm\pip$, 
$\Bz\to\DDstarm\rhop$, $\Bz\to\DDstarm\aonep$ and $\Bz\to\jpsi\Kstarz$. 
A variety of \Dzb, \Dm, and \Dstarm\ modes are used to achieve reasonable efficiency
despite the typically small branching fractions for any given \B\ decay channel.

\subsubsection{Event selection}


We select \Dzb, \Dm\ and $\Dstarm$ mesons with the following
criteria. \Dzb\ candidates are identified in the 
decays channels \Kp \pim, \Kp \pim \piz, \Kp \pip \pim \pim\ and \KS
\pip \pim. \Dm\ candidates are selected using the \Kp \pim \pim\
and \KS \pim\ modes. The $\Dstarm$ candidates are found using the decay 
$\Dstarm\to\Dzb\pim$. 
Charged and neutral kaons are required to  have a momentum
greater than 200\mevc. The same criterion was applied to the pion in 
$\Bz\to\DDstarm\pip$, $\Bz\to\DDstarm\rhop$ decay.  
For the decay modes $\Bz\to\DDstarm\aonep$, the pions are  required to have 
momentum larger that  150\mevc\ . 
\Dzb\ and \Dm\ candidates are required to lie within
$\pm 3\sigma$, calculated on an event-per-event basis, of the nominal
masses. Pull distributions for the \Dzb\ and \Dm\ meson masses  have been measured 
in data and were found to have rms in the range of 1.1--1.2 when fitted to a Gaussian form. 
For $\Dzb\to\Kp \pim \piz$, we only reconstruct the dominant resonant
mode $\Dzb\to\Kp \rhom$, followed by $\rhom\to\pim \piz$. The \pim \piz\ mass is
required to lie within $\pm 150$\mevcc\ of the nominal $\rho$ mass and
the angle between the \pim\ and \Dzb\ in the $\rho$ rest frame, 
$\theta^*_{\Dz\pi}$, must satisfy $\mid \cos \theta^*_{\Dz \pi} \mid > 0.4$.
Finally, all \Dzb\ and \Dm\ candidates are required to  have a momentum greater than 1.3\gevc\
in the \FourS\ frame and a $\chi^2$ probability of the topological vertex fit 
greater than 0.1\%.
A mass constrained fit is applied to candidates satisfying aforementioned requirements
and is used in the subsequent reconstruction chain.

We form \Dstarm\ candidates by combining a \Dzb\ with a pion which has
momentum greater than 70\mevc. 
The soft pion is constrained to originate from the beamspot
when the \Dstarm\ vertex is computed~\cite{BabarPub0017}. 
To account for the small energy release in the 
decay $\FourS \to \B\Bbar$ (resulting in a small transverse flight of
the \B\ candidates), the assumed vertical size of the beam spot is inflated to  
40\mum. Monte Carlo simulation was used  to verify that this inflation 
does not introduce any significant bias in the selection or in the
$\Delta t$ measurement.
\Dstarm\ candidates are then required to have $\Delta m=m(\Dzb\pim)-m(\Dzb)$
within $\pm 1.1$\mevcc\ of the nominal value
for $\Dzb\to\Kp \pim \piz$ mode and $\pm 0.8$\mevcc\ for all the
other modes. This corresponds to about
$\pm 2.5\sigma$, where the resolution is estimated by taking a weighted
average of the core and broad Gaussian components of the observed $\Delta m$ distributions.

\Bz\ candidates are formed by combining a \Dstarm, \Dm\ or \jpsi\ candidate
with a \pip, \rhop, \aonep\ or \Kstarz.  For $\Bz\to\Dstarm
\rhop$, the \piz\ from the \rhop\ decay is required to have an
energy greater than 300\mev. For $\Bz\to\Dstarm\aonep$,
the \aonep\ is reconstructed by combining three charged
pions, with invariant mass in the range 1.0 to 1.6\gevcc. In
addition, the $\chi^2$ probability of the vertex fit of the \aonep\ candidate is required to be
greater than 0.1\%.

As described in Section \ref{sec:PID}, kaon identification is used to
reject background. For most \Bz\ modes, no identification or only a loose selection
is enough to achieve signal purities of 90\%. 
In the $\Bz\to\Dm\aonep$ mode reconstruction, a  tight kaon identification is required to alleviate 
large combinatorial backgrounds.



 In order to suppress continuum background, in addition to the $R_2$ requirement, 
we calculate the 
thrust angle, $\theta_{th}$, 
defined as the angle between the thrust axis of the particles which form the
reconstructed \Bz\ candidate and the thrust axis of the remaining
tracks and unmatched clusters in the event, computed in the \FourS\ frame. 
The two thrust axes are almost uncorrelated in \BB\ events, 
because the \Bz\ mesons are almost at rest in the \FourS\ rest frame.
In continuum events, which are more jet-like,
the two thrust axes tend to have small opening angles.
For final states with a \Dstar\ and 2 (3) pions we require
$|\cos\theta_{th}|<0.9$ (0.8) for the $\Dzb\to\Kp\pim$ and \Kp \pim \piz\
modes and 0.8 (0.7) for the  $\Dzb\to\Kp \pip \pim \pim$ and \KS \pip
\pim. No such requirement is applied in the case of the $\Bz \to \Dstarm \pip$ mode.
In modes which contain a \D\ and a single (2,3) pion(s) in the final
state, we require $|\cos\theta_{th}|<0.9$ (0.8, 0.7). 

The $\Bz\to\jpsi \Kstarz$ selection is the same as described in Ref.~\cite{BabarPub0005}.
The $\jpsi\to\epem$ candidates are formed from pairs of oppositely-charged tracks assumed to be 
electrons, as explained in Section~\ref{sec:PID}.    
At least one of the decay products must be positively identified as an electron or,
if outside the acceptance of calorimeter, must be consistent with the electron hypothesis from the 
\dedx\ measurement in the drift chamber.
If both tracks are in the calorimeter acceptance and have a 
value of $E/p$ larger than 0.5, an algorithm for the recovery of
Bremsstrahlung photons~\cite{BabarPub0005}  is used.
The $\jpsi\to \mu^+ \mu^-$ candidates are formed from oppositely-charged tracks
identified as  
muons, as described in Section~\ref{sec:PID}.  At least one of the decay products must be positively identified as a muon,
and the other, if in the acceptance of the calorimeter, must be consistent with a 
minimum ionizing particle. 
We retain only those $\jpsi$ candidates with an invariant mass in the range 
$2.95<m(\jpsi)<3.14$\gevcc\ for
the \epem\ mode and $3.06<m(\jpsi)<3.14$\gevcc\ for the $\mu^+\mu^-$ mode. We reject events with 
$\cos\theta_{th}$ greater than 0.9.

\Bz\ candidates are characterized by a pair of nearly uncorrelated 
kinematic variables, the 
difference between the energy of the \Bz\ candidate and the 
beam energy in the $\Upsilon(4S)$ center-of-mass frame, $\Delta E$, 
and the beam energy substituted mass, \mes~\cite{BabarPub0017}.
In the $(\mes,\deltae)$ plane the signal region is defined 
as $5.270 < \mes <  5.290$\gevcc\ and $|\Delta E|< 3\sigma_{\deltae}$. 
The sideband region is defined as $5.2 < \mes < 5.26$\gevcc\ and 
$|\Delta E|< 3\sigma_{\deltae}$. 
The value of $\sigma_{\deltae}$, the $\Delta$E resolution, is mode-dependent and \
varies between 19 to 40 \mev.
When multiple \Bz\ candidates  (with $\mes > 5.2$\mevcc) are found in the same event, 
the candidate with the smallest value of $|\deltae|$ is selected.


\subsection{Sample composition}
We use the \mes\ sideband region  to
measure the combinatorial background fraction, to characterize the background
$\Delta t$ distribution, and to measure the fraction of mixed events
contributed from background under the \Bz\ signal peak. It is therefore
useful to check that the background composition in the \mes\
sideband region is similar to that in the signal region.
The validity of this  background estimation procedure was checked on 
2.0 fb$^{-1}$ of simulated data. 
We also compared the \mes\ sideband shape in data with the Monte
Carlo and found good agreement. 

The \Bz\ signal yield and sample purity extracted from 
fits to the \mes\ distribution are summarized in 
Table~\ref{tab:HadronicB0Yield}. 
The net \Bz\ signal sample, before applying any decay vertex requirements, 
consists of $2577\xpm\ 59$ signal candidates with a purity of about 86\%.


\begin{table}[!htb]
\caption{Two-body hadronic \Bz\ decay candidate yields and signal
purities from the fit to the \mes\ distribution. Signal purities are
estimated for $\mes> 5.27$\gevcc.}
\vspace{0.3cm}
\begin{center}
\begin{tabular}{|l|c|c|} \hline
  Decay mode   & Number of \Bz\ candidates  & S/(S+B) [\%] \\
\hline\hline
 $\Bz\to\Dstarm \pip$    & $622\pm 27$  & 90  \\ 
 $\Bz\to\Dstarm \rhop$   & $419\pm 25$  & 84  \\ 
 $\Bz\to\Dstarm \aonep$  & $239\pm 19$  & 79 \\ 
 $\Bz\to\Dm \pip$        & $630\pm 26$  & 90 \\  
 $\Bz\to\Dm \rhop$       & $315\pm 20$  & 84  \\ 
 $\Bz\to\Dm \aonep$      & $225\pm 20$  & 74  \\ 
 $\Bz\to\jpsi \Kstarz$   & $194\pm 15$  & 90 \\ 
\hline
  Total                  & $2577\pm 59$ & 86  \\ 
\hline
\end{tabular}
\end{center}
\label{tab:HadronicB0Yield}
\end{table}

In order to select \Bz\ candidates with well understood \B\ decay vertex error, 
we require the convergence of the topological fit for \Bz\  decay vertex. \Bz\ 
candidates with $|\Delta z| > 3.0$\mm\ or with $\sigma_{\Delta z}>400$\mum\ are 
removed. 

\begin{figure}[tbhp]
\begin{center}
\begin{tabular}{lr}
\mbox{\epsfxsize=7.5cm\epsffile{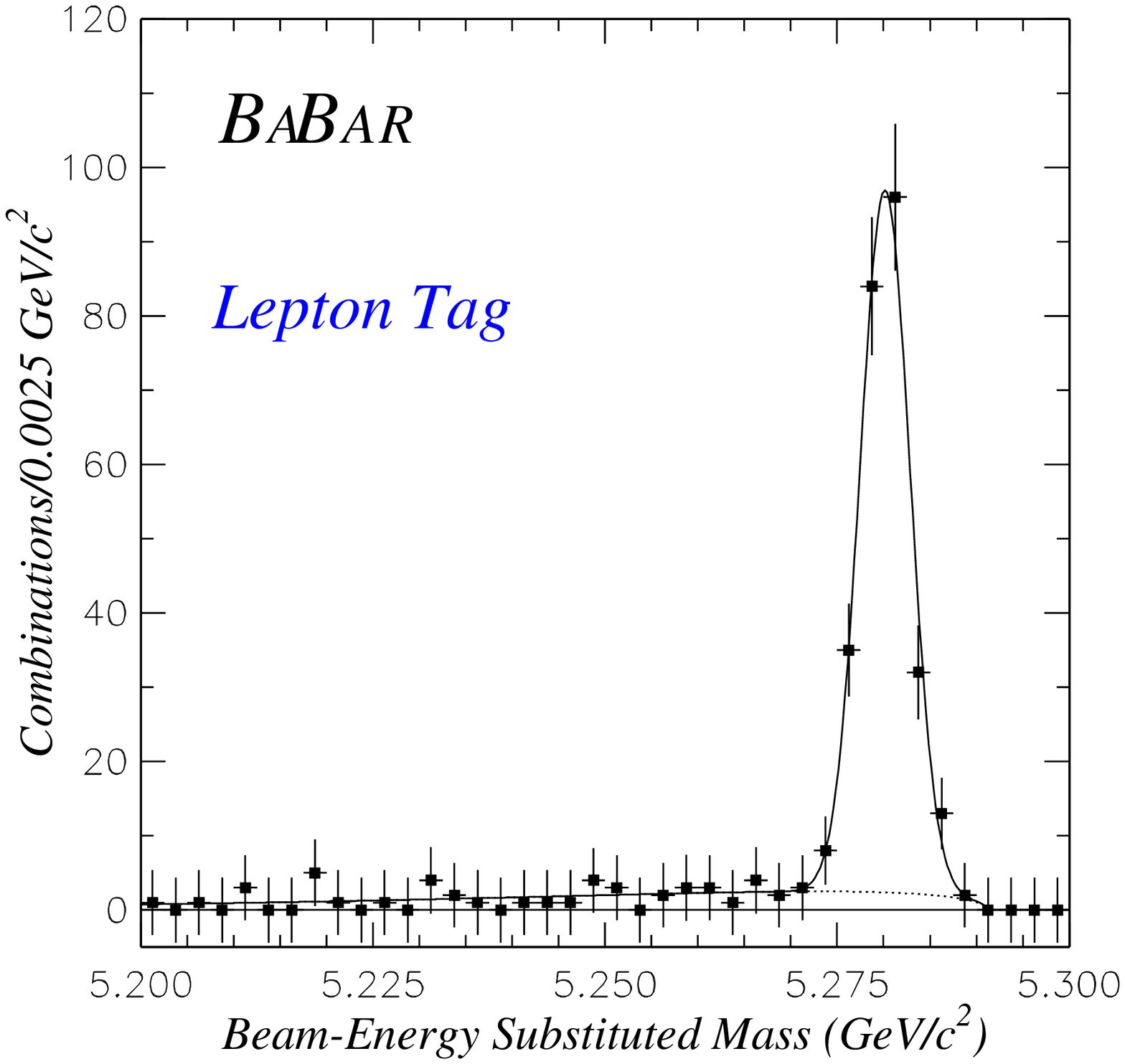}} &
\mbox{\epsfxsize=7.5cm\epsffile{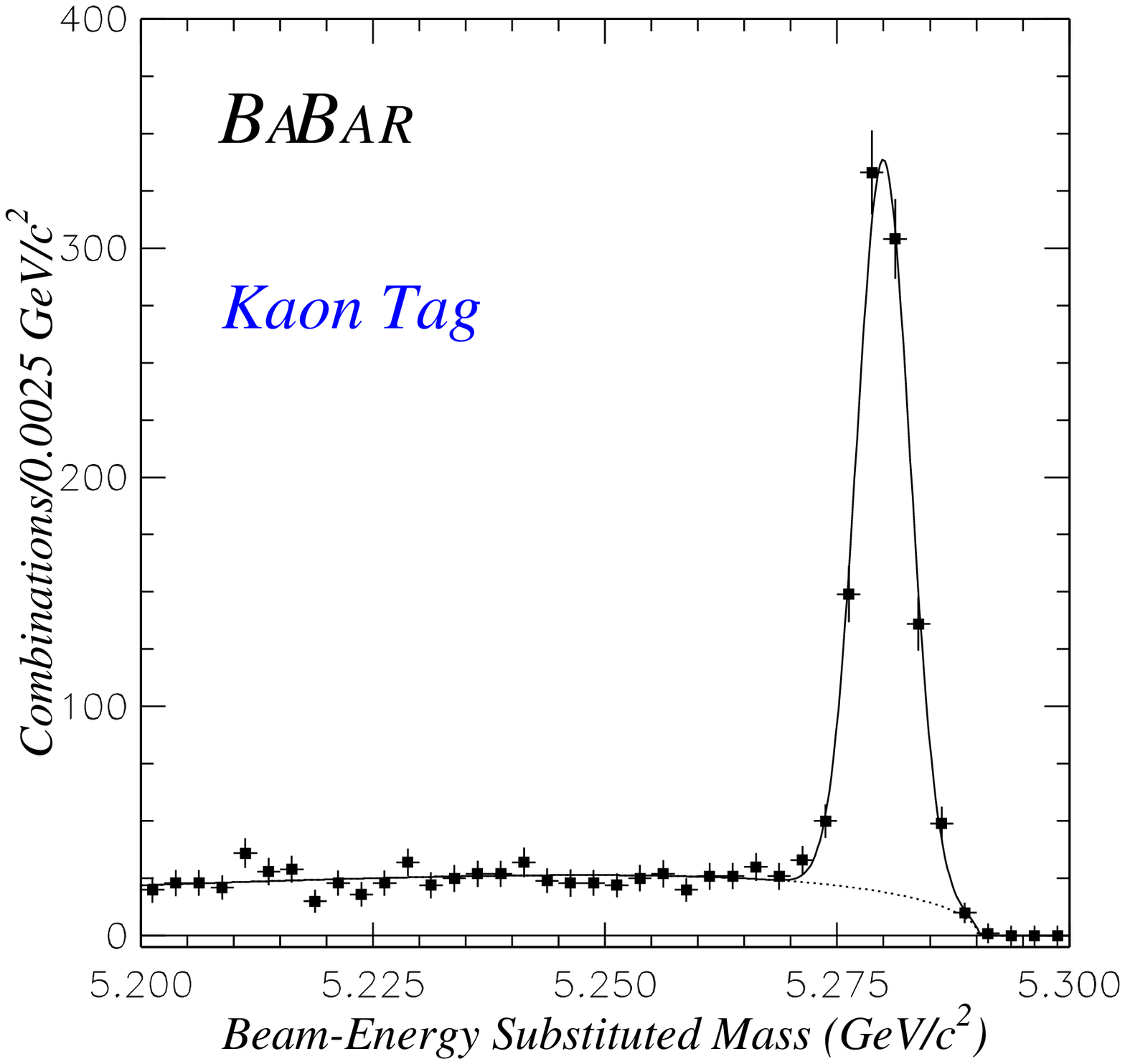}} \\
\mbox{\epsfxsize=7.5cm\epsffile{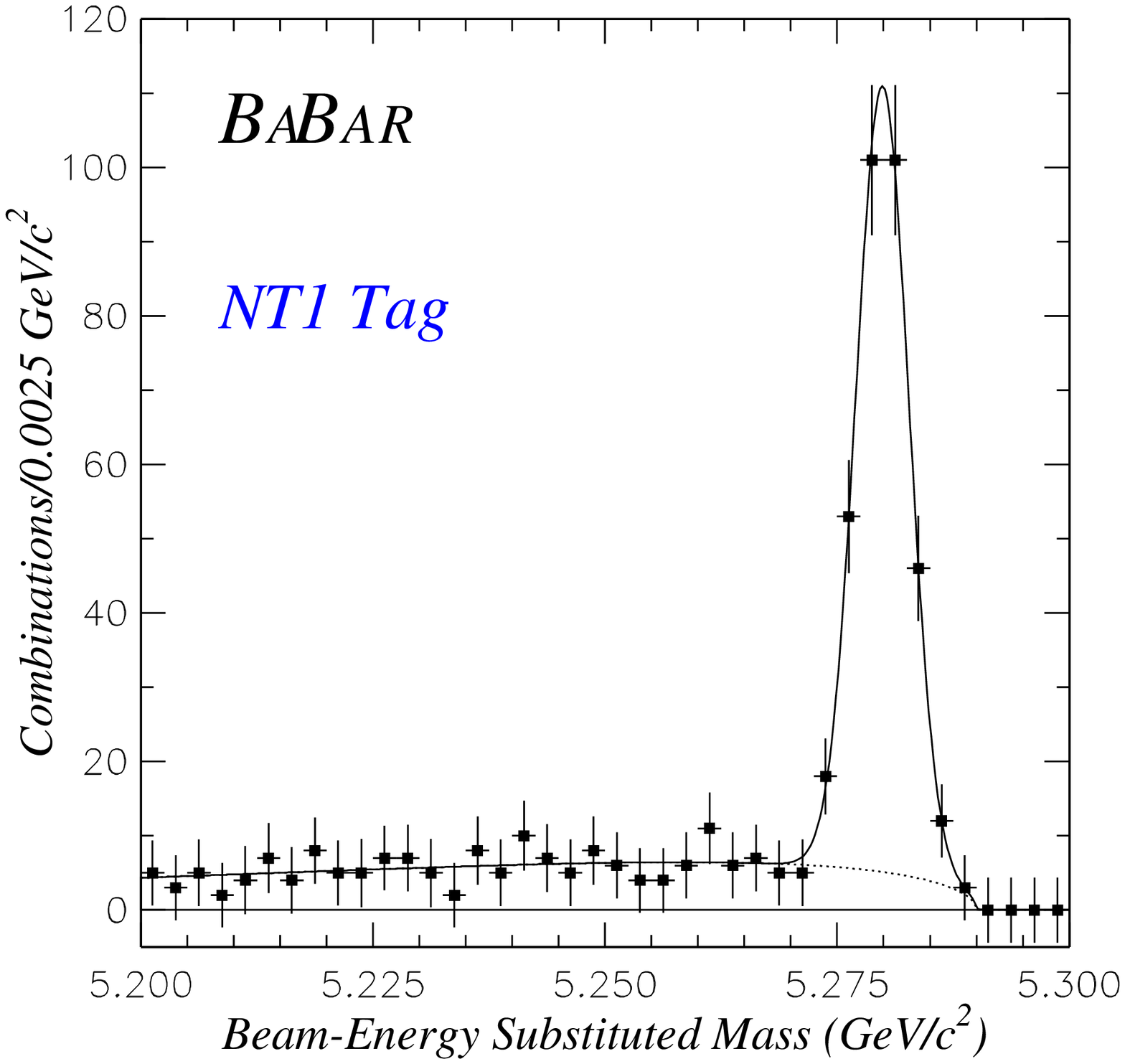}} &
\mbox{\epsfxsize=7.5cm\epsffile{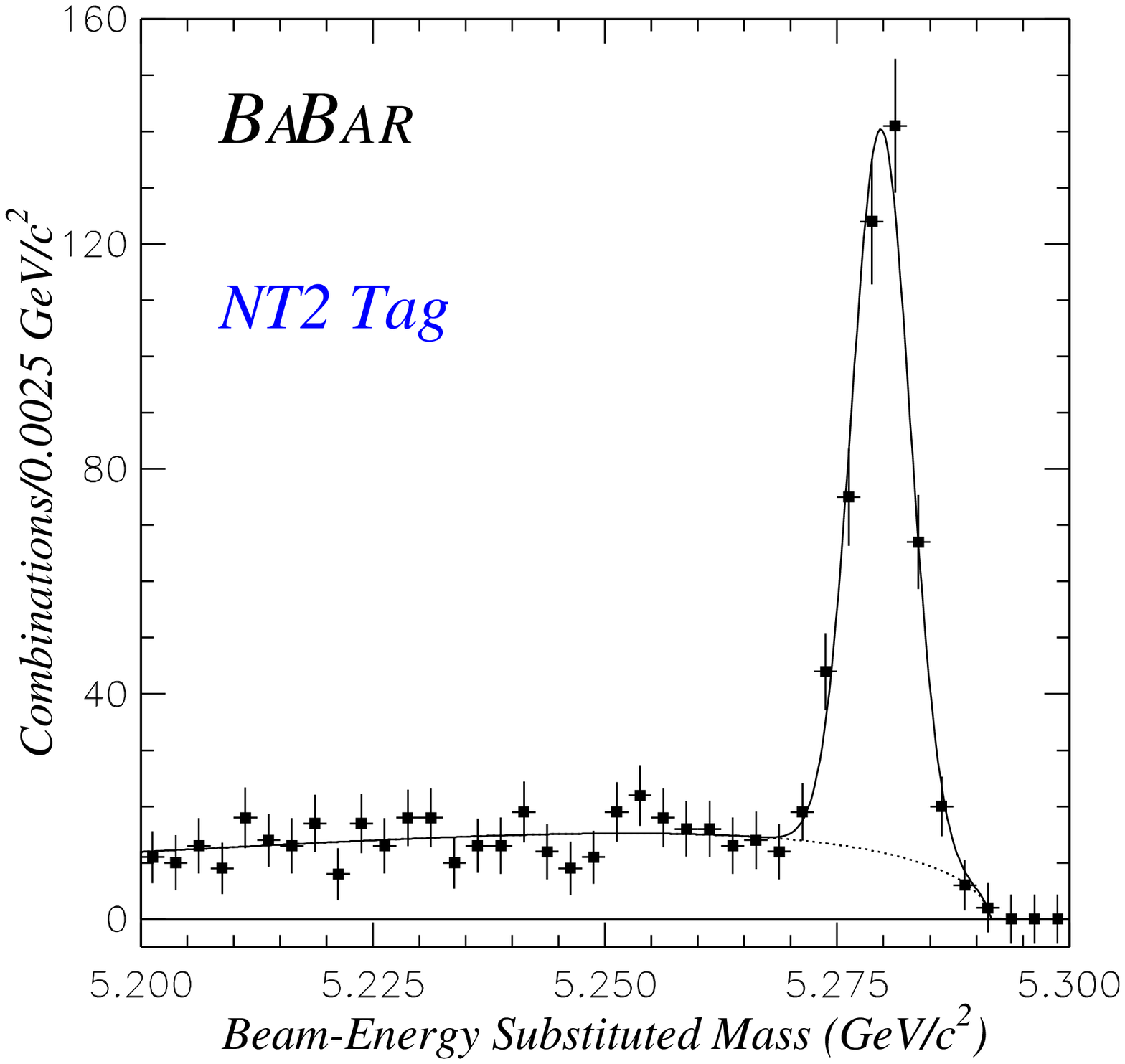}}  \\
\end{tabular}
\end{center}
\caption{\mes\ distribution for each tagging category ({\tt
Lepton}, {\tt Kaon}, {\tt NT1} and {\tt NT2}) for all the hadronic
\Bz~modes.
\label{fig:b0mix.excl.mb-cats}}
\end{figure}

The efficiency for each tagging 
category is calculated as the number  of signal events for
each tag, divided by the total number of signal events after vertex cuts are imposed.
The tagging efficiency and signal purity for the
individual tagging categories in data are extracted 
from fits to the \mes\ distributions shown in
Fig.~\ref{fig:b0mix.excl.mb-cats} and are listed in  
Table~\ref{tab:tagging-excl}. The distributions are
fitted with a Gaussian distribution for the
signal and the ARGUS background function  for the background 
parametrization~\cite{ARGUS_bkgd}. All fits have good confidence levels.

\begin{table}[!htb]
\begin{center}
  \caption{Tagging efficiencies for hadronic \Bz\ decays and signal purities 
in data separately for the four tagging categories. Signal purities are
estimated for $\mes > 5.27$\gevcc.} \vspace{0.3cm}
\label{tab:tagging-excl} 
  \begin{tabular}{|l|ccc|}\hline
Tagging Category & {Efficiency [\%]} & {$B$ candidates}  & S/(S+B) [\%]\\
    \hline\hline
{\tt Lepton}&   10.5 \xpm\ 0.6 &  260 \xpm\ 17 & 95  \\ 
{\tt Kaon}  &   36.7 \xpm\ 1.0 &  918 \xpm\ 34 & 86  \\
{\tt NT1}   &   12.0 \xpm\ 0.7 &  305 \xpm\ 19 & 89  \\
{\tt NT2}   &   16.4 \xpm\ 0.7 &  405 \xpm\ 24 & 81  \\ \hline
All tags    &   75.6 \xpm\ 0.9 & 1886 \xpm\ 49 & 87  \\ 
    \hline
  \end{tabular}
\end{center}
\end{table}

\subsection{Semileptonic \boldmath $B$ decays}
\label{subsec:Sample_semileptonic}
The semileptonic decay $\Bz\rightarrow D^{*-} \ell^+ \nu$,
with a measured branching fraction of $4.6 \pm 0.27 \%$~\cite{PDG},  
is a copious source of \Bz\ mesons.  We reconstruct the
$D^{*-}$ through its decay to $\Dzb\pi^-$, and use the three 
\Dzb\ decay modes
$K^+\pi^-$, $K^+\pi^+\pi^-\pi^-$  and $K^+\pi^-\pi^0$.   

\subsubsection{Event selection}

We reconstruct $\bar D^0$ candidates in the three modes listed above.
All reconstructed $\bar D^0$ candidates are required to  have an invariant mass within
$\pm 2.5 \sigma$ of the nominal $D^0$ mass.  
The $\bar D^0$ topological vertex fit  is required to  have a $\chi^2$
probability greater than 1\%; for the $\bar D^0\rightarrow K^+\pi^-\pi^0$
mode this vertex fit includes a kinematic constraint on the $\pi^0$ mass.
There are no additional requirements for $\bar D^0\rightarrow K^+\pi^-$.
For $\bar D^0\rightarrow K^+\pi^+\pi^-\pi^-$ and $\bar D^0\rightarrow K^+\pi^-\pi^0$
we require loose $K$ and $\pi$ particle identification as described
in Section~\ref{sec:PID}, and a minimum $\pi^0$ momentum of 200\mevc.
In addition, the $K$ and $\pi$ candidates  are required to  have
momenta greater than 200 and 150\mevc, respectively, for the mode
$\bar D^0\rightarrow K^+\pi^+\pi^-\pi^-$.  The decay  
$\bar D^0\rightarrow K^+\pi^-\pi^0$ occurs mostly through resonant substructures.
The $\rho$ and $K^*$ resonances dominate and we use the measured Dalitz weights
to construct an event-by-event probability and select events using this quantity
to suppress combinatorial background\footnote[2]{ The Dalitz weights are used in the 
time-integrated analysis. The time-dependent analysis uses the  
$\bar D^0\rightarrow K^+\rho^-$ sample only.}.

$\bar D^0$ candidates satisfying the above requirements are combined with
all charged tracks, having a minimum transverse momentum of 50\mevc\ and
charge opposite to that of the candidate kaon, to form $D^*$ candidates.
The mass difference $m(D^{*-}) - m(\bar D^0)$ is 
calculated and required to lie within 2.5$\sigma$ of the nominal value.

Finally, $D^*$ candidates are combined with electron or muon 
candidates, with momentum greater than 1.2\gevc\
and satisfying the lepton
identification requirements described in Section~\ref{sec:PID}.
The $D^{*}$ -- lepton topological vertex 
fit is required to  have a $\chi^2$ probability greater than 1\%, and the lepton
and $D^{*}$ candidates must have opposite charge.
The $D^{*}$ and lepton tend to be back-to-back in the
\Bz\ rest frame, so we require 
$\cos\theta (D^* - l) < 0$ where $\theta(D^*-l)$ is the angle between
the $D^{*}$ and the lepton in the center-of-mass frame.

The neutrino cannot be reconstructed, but
we require that the candidate
four-momenta be consistent with a missing particle of zero mass:
\begin{equation}
(p_B - p_{D^*} - p_l)^2 = p_{\nu}^2 = 0 .
\end{equation}
\noindent
Solving this equation in the \FourS\ frame, we obtain a constraint on the angle between the
\Bz\ and the $D^*-l$ system:
\begin{equation}
\cos\theta (B, D^*l)  = {{M_B^2 + M_{D^*l}^2 - 2E_BE_{D^*l}}\over
                          {2|\vec p_B| |\vec p_{D*l}|}} .
\end{equation}
\noindent
Only the mass and scalar momentum of the initial state \Bz, both of which are known, is required
to calculate the angle $\theta (B, D^*l)$ and not the flight direction.
In  $B^0\rightarrow D^* l\nu$ decay,
the cosine of this angle must lie in the region $(-1,+1)$.
Allowing for detector resolution effects in the reconstructed
momenta and angles,  we require
$|\cos\theta(B-D^*l)| < 1.1$.

After applying these criteria, we obtain a sample of $7517 \pm 104$
\btodstarlnu\ events: $3101\pm 64$ in the $\bar D^0\rightarrow K^+\pi^-$ mode,
$1986 \pm 51$  in the $\bar D^0\rightarrow K^+\pi^-\pi^0$ mode, and
$2430 \pm 56$  in the $\bar D^0\rightarrow K^+\pi^+\pi^-\pi^-$ mode.

\subsubsection{Sample composition}

\begin{figure}[tbhp]
\begin{center}
\begin{tabular}{lr}
\mbox{\epsfxsize=7.5cm\epsffile{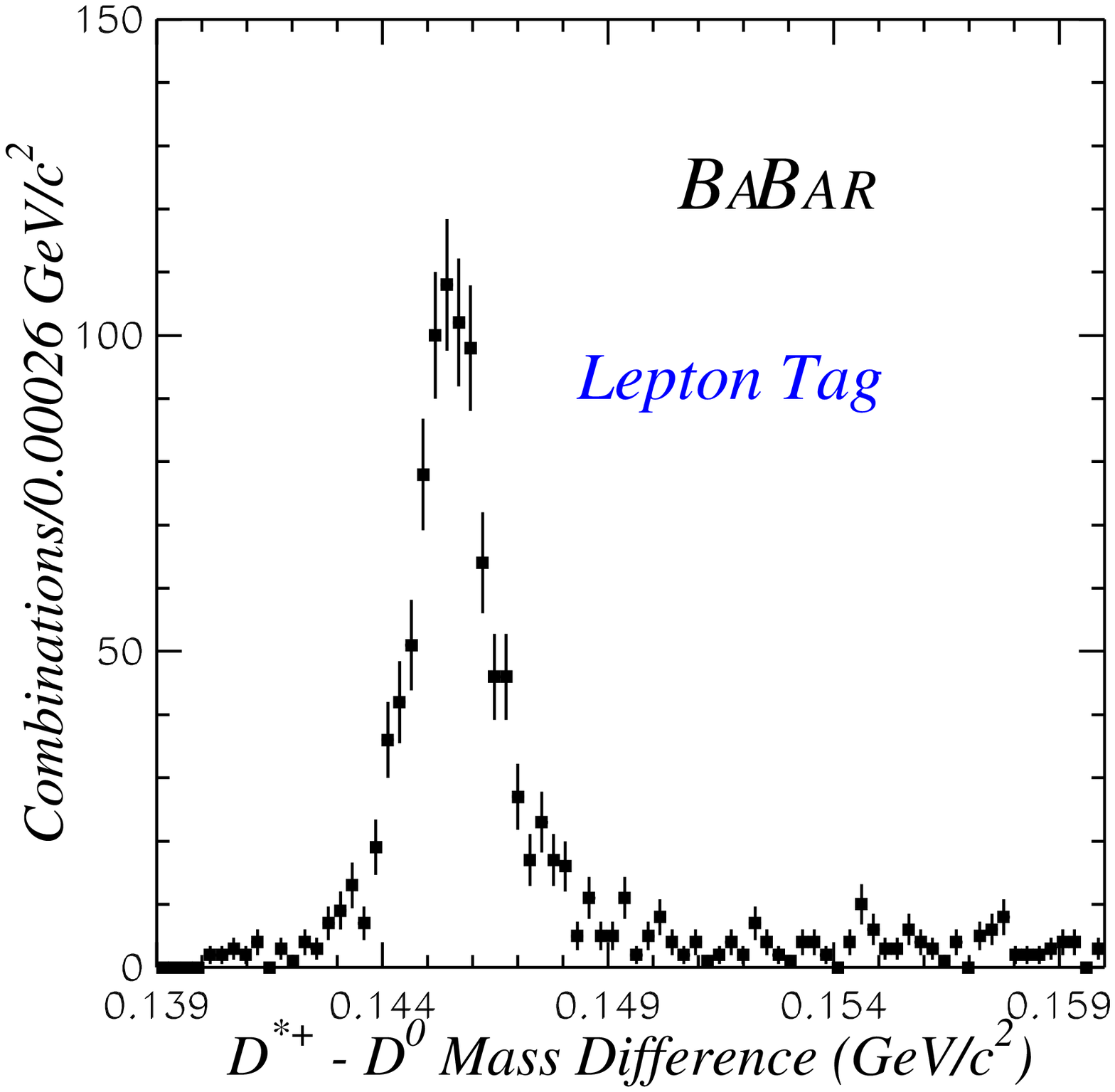}} &
\mbox{\epsfxsize=7.5cm\epsffile{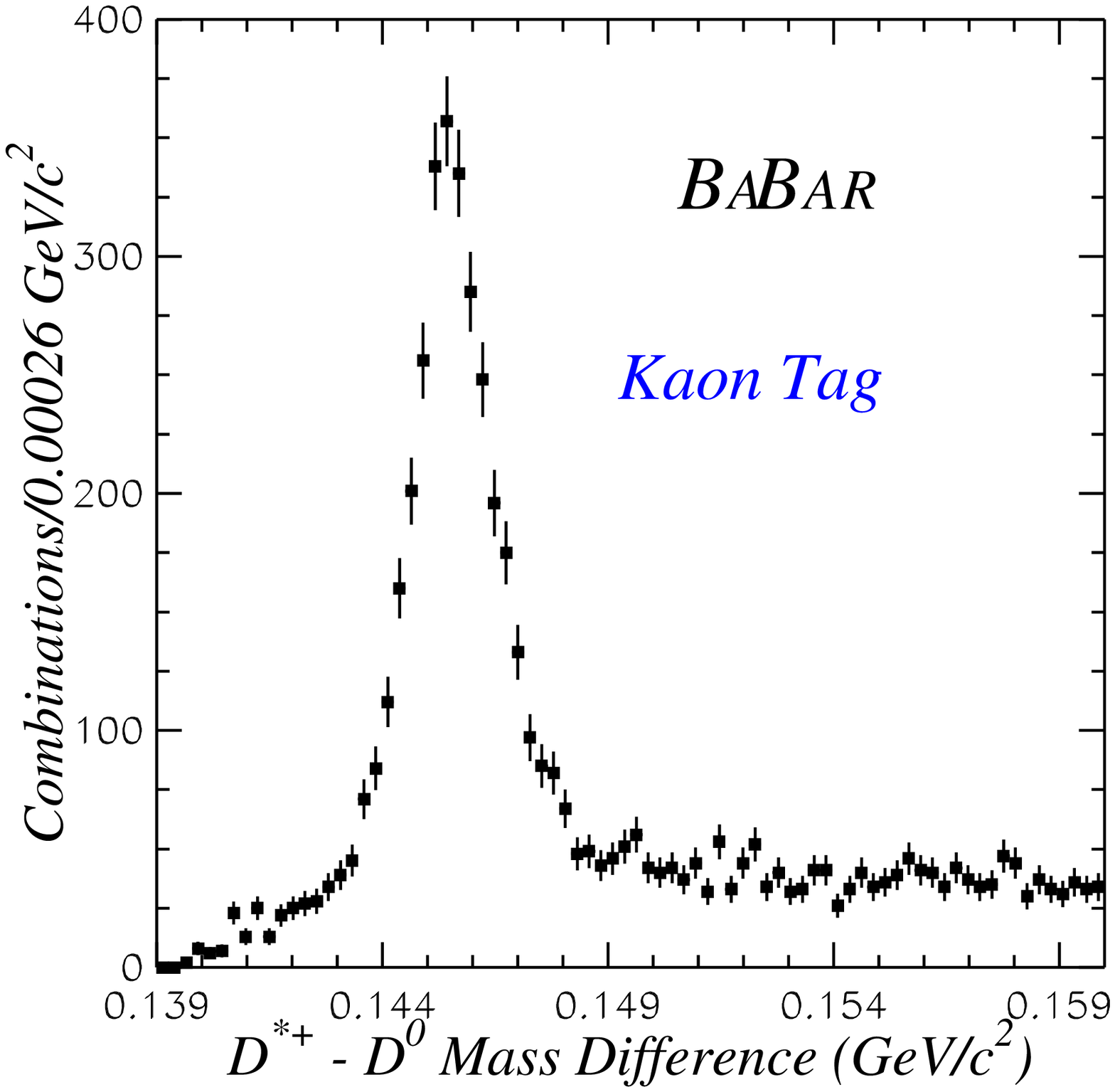}} \\
\mbox{\epsfxsize=7.5cm\epsffile{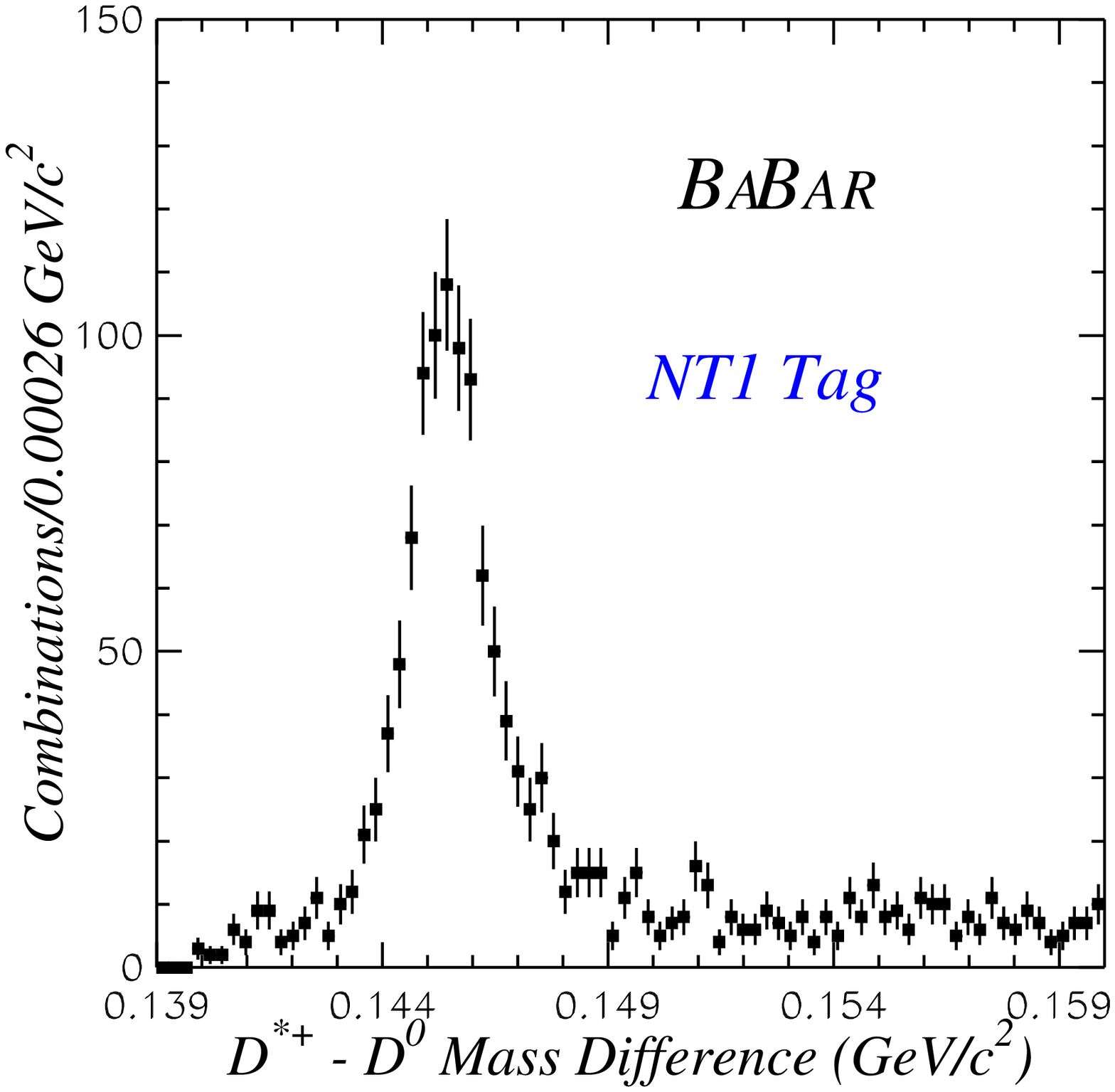}} &
\mbox{\epsfxsize=7.5cm\epsffile{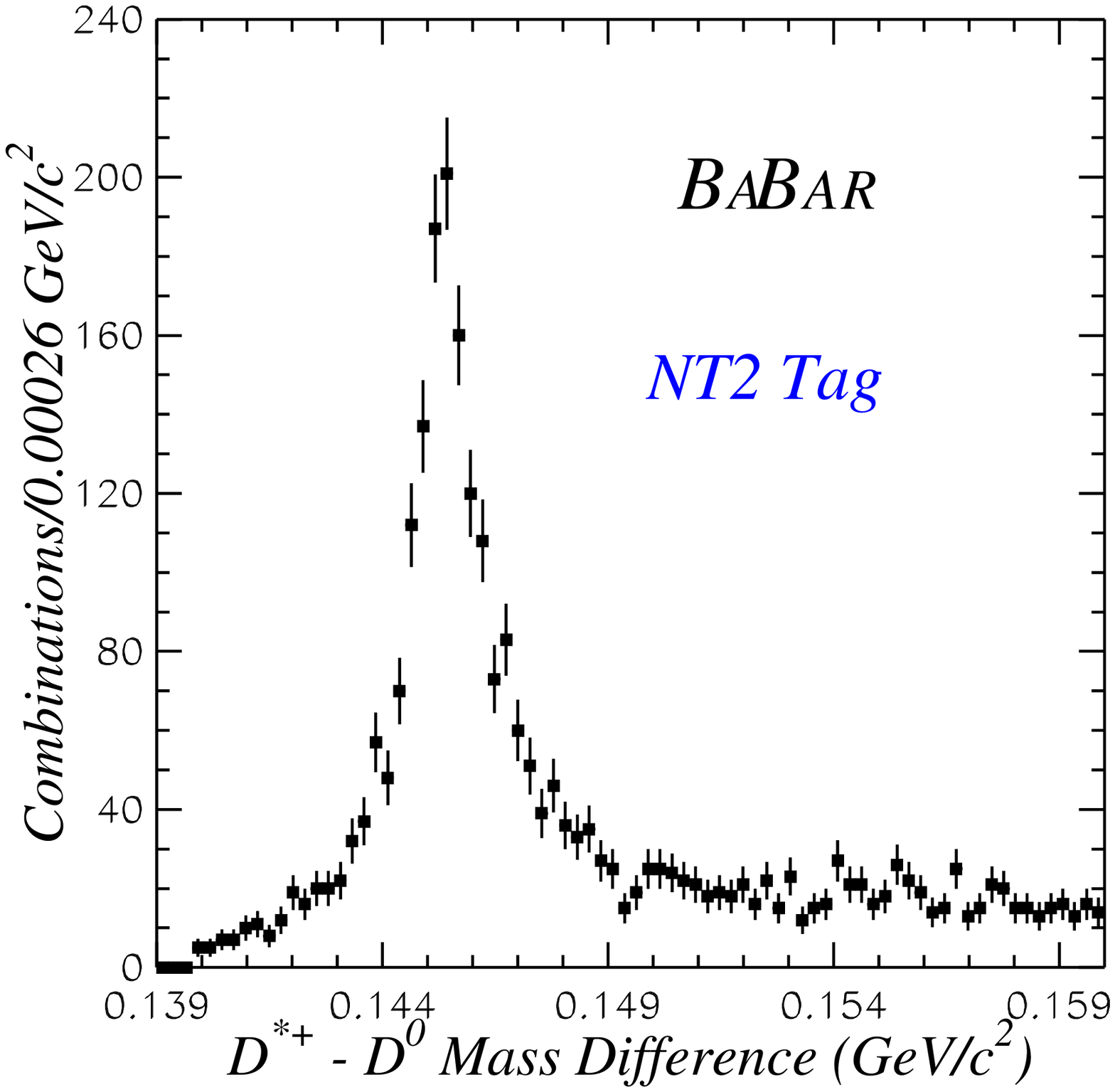}}  \\
\end{tabular}
\end{center}
\caption{ $D^*~-D^0$ mass difference distribution  for each tagging category ({\tt
Lepton}, {\tt Kaon}, {\tt NT1} and {\tt NT2}) for the $D^* l\nu$ sample.
\label{fig:b0mix.dstarlnu-cats}}
\end{figure}

The events are flavor tagged as described in Section~\ref{sec:Tagging}.
The $D^*-D^0$ mass difference distributions for each tagging category are 
shown in Fig.~\ref{fig:b0mix.dstarlnu-cats}.
The backgrounds are evaluated separately for each tag category.
Backgrounds are larger for the semileptonic modes than for the
hadronic modes and originate from a variety of sources.
Each source of background is evaluated using
a control sample that is taken from data whenever possible.  
The background control
samples are used to measure the background fractions, to
characterize the background $\Delta t$ distribution, and to measure
the fraction of mixed events contributed by each source of background.

We divide the backgrounds to $B^0 \rightarrow D^* l\nu$ into three
types:  events with an incorrectly reconstructed $D^*$ 
(``combinatorial'' background), events in which a true $D^*$ 
is combined with an incorrect lepton candidate (``wrong-lepton'' background),
and events in which a correctly identified
$D^*$-lepton pair originates from semileptonic 
$B^+$ decay, for example from the decay $B^+\rightarrow \bar{D}^{**0}l^+\nu$,
followed by $\bar{D}^{**0}\rightarrow D^{*-}n\pi^+$ ( ``$B^+$ background'').
The symbol $\bar{D}^{**0}$ refers to an admixture of orbitally or radially excited 
charmed meson resonances  which decay via strong interaction into a $D^{*-}$. 
Events of the type $B^0\rightarrow D^{**-}l^+\nu$ are considered
as signal. 

\bigskip\bigskip
{\noindent\bf Combinatorial background}
\medskip

The fraction of combinatorial background, due to
falsely reconstructed $D^*$ candidates, is estimated by fitting the 
$\Delta m (D^* - D^0)$ distributions.
We use a Gaussian distribution to characterize the signal and a
threshold function with a sharp rise followed by an exponential tailoff 
to characterize the background shape. 
the signal region is defined to lie within $\pm 2.5 \sigma$ of the peak in
$\Delta m (D^* - D^0)$, while events in the
the sideband region
$0.150 < \Delta m(D^*-D^0) < 0.160$\gevcc\ provide the
combinatorial background control sample.

\bigskip
\bigskip
{\noindent\bf Wrong-lepton background}
\medskip

There are four potential sources of background events in which a
real $D^*$ is combined with a wrong lepton.  We consider each
in turn.

First, there are events in which the lepton candidate is not
a real lepton, but is misidentified as such.  
This ``fake lepton'' background is estimated
from data by selecting
events in which a track candidate that has failed very
loose lepton criteria is substituted for the lepton candidate.
These events are weighted with the
lepton mis-identification probabilities 
measured in  data to
estimate the fraction of fake lepton background, after subtraction
of the combinatorial background contribution.  

The second type of wrong lepton events is due to a real $D^*$
from one $B$ combined with a real lepton from the other $B$.
These ``uncorrelated lepton'' backgrounds
are due  to \BzBzb\ events where mixing
has occurred, yielding the right $D^* l$ sign combination, or
events in which a lepton originating from a secondary charm
decay is combined with a $D^*$ from the other $B$.
The $D^*$ and lepton directions are uncorrelated in these events,
prior to application of the $\cos\theta(D^* l)$ and $\cos\theta(B-D^*l)$ 
selection criteria.  

To estimate this background from data,
the lepton momentum vector in the \FourS\ center-of-mass frame is 
parity-inverted, prior to the
calculation of the $\theta(D^*-l)$ and $\theta(B-D^*l)$. Events
with a correlated back-to-back $D^*-l$ pair fail these
criteria after the lepton momentum is flipped, but events with
a randomly correlated  $D^*-l$ pair pass with approximately
the same efficiency as in the original sample.  After removing the
remaining signal contribution, and correcting for residual 
combinatorial background, we estimate the uncorrelated lepton
background from the flipped-lepton control sample.

Third, there are decays of the type $B^0\rightarrow D^* D X$, where
the $D$ decays semileptonically producing  a non-primary lepton.
The momentum requirement on the lepton rejects most of these 
``cascade lepton'' events;
the remaining fraction, estimated from Monte Carlo simulation, is
less than 1\% and has been neglected.

Finally, $c\bar c$ events can produce a real $D^*$ and a real lepton
in a back-to-back configuration.  The $c\bar c$ background fraction 
has been estimated using combinatorial-subtracted off-resonance data.  
We use the off-resonance data as a control sample to
characterize the $\Delta t$ distribution and the mixed event
contribution from this background source.

\bigskip
{\noindent\bf \boldmath $B^+$ background}
\medskip

In addition to the well-studied semileptonic decays
$B\rightarrow Dl\nu$ and $B\rightarrow D^*l\nu$, a significant fraction
of $B$ semileptonic decays involve additional final state particles,
either produced through $D^{**}$ resonances or non-resonantly.
These processes of the type $B\rightarrow D^* (n\pi)l\nu$ contribute
to both the neutral and charged semileptonic $B$ decays.

For the purposes of this analysis, neutral semileptonic decays
$B^0\rightarrow D^{*} (n\pi) l\nu$ that pass our event selection
criteria are considered to be part of the signal.  They contribute
equally to the measurement of $\Delta m_d$, and the additional
low-momentum pion does not affect the tagging algorithm.

However, the charged $B$ decays of the type 
$B^-\rightarrow D^{*+}(n\pi) l^- \nu$ are considered as background for this
analysis.  They do not oscillate and must be corrected for in
extracting $\Delta m_d$; their mistag rate may differ from that of
$B^0$ decays as well.

We assume a lifetime and 
$\Delta t$ resolution for charged $B$ events 
that is the same as that for the $B^0$ signal events.
The mistag rate is estimated directly
from data using fully reconstructed $B^+$ decays.  
It is difficult to accurately estimate the fraction of this
background, since it cannot be cleanly separating from the signal.
To estimate the fraction of events due to
charged $B$ decay, we rely on an estimate
of their production rate, combined with a Monte Carlo study of
the event selection efficiency for these modes.

The inclusive branching fraction, 
${\cal B}(B^+\rightarrow D^{*-} (n\pi) l^+\nu)=1.25\xpm 0.16\%$ 
is taken from LEP measurements~\cite{Aleph}. 
We use Monte Carlo simulation to study the efficiency of the
event selection criteria for events of this type, averaging over
several resonant $D^{**}$ and non-resonant states.
The average efficiency is found to be $4.25\xpm\ 3.0\%$,
where a conservative systematic error has been assigned due to the 
lack of knowledge of the relative decay fractions for the possible
modes.  The product of the branching ratio times the 
event selection
efficiency, ${\cal B}\times \epsilon(B^+)$, 
and the corresponding product ${\cal B}\times \epsilon(B^0)$ for the 
neutral $B$ signal events are computed, where
decays of the type $\Bzb\rightarrow
D^{*+}(n\pi)l^-\nu$ are included as part of the signal. By this method, we find:
\begin{equation}
f(B^+) = {{{\cal B}*\epsilon(B^+)}\over
{{\cal B}*\epsilon(B^+)+ {\cal B}*\epsilon(B^0)}} = 7.1\xpm\ 5.0\%
\end{equation}


\bigskip
{\noindent\bf Background summary}
\medskip

The background contributions in \btodstarlnu\ events,
averaged over $D^0$ decay modes, are summarized by tag category 
in Table~\ref{tab:D*lnu_bkgd}.  
\begin{table}
\begin{center}
\caption{Sample composition and yields in tagged 
$B^0\rightarrow D^* l\nu$ events} \vspace{0.3cm}
\label{tab:D*lnu_bkgd}
\begin{tabular}{|l|c|c|c|c|}
\hline 
                & Lepton              & Kaon            & {NT1}            & {NT2}  \\ \hline
\hline
Combinatorial   &  0.063\xpm 0.008   & 0.161\xpm 0.007  & 0.128\xpm 0.011  & 0.158\xpm 0.009   \\ 
Fake Lepton     &  0.032\xpm 0.007   & 0.035\xpm 0.007  & 0.031\xpm 0.007  & 0.030\xpm 0.007   \\ 
Uncorr. Lepton  &  0.015\xpm 0.016   & 0.035\xpm 0.012  & 0.024\xpm 0.013  & 0.028\xpm 0.013   \\ 
$c\bar c$       &  0.000\xpm 0.007   & 0.026\xpm 0.010  & 0.025\xpm 0.017  & 0.056\xpm 0.021   \\ 
$B^{\pm}$       &  0.062\xpm 0.040   & 0.052\xpm 0.031  & 0.055\xpm 0.038  & 0.051\xpm 0.034   \\  
Signal Fraction &  0.827\xpm 0.040   & 0.691\xpm 0.031  & 0.737\xpm 0.038  & 0.677\xpm 0.034    \\
 \hline
Tagged Event Yield & 863 \xpm 32  &  2804 \xpm 63  &  850 \xpm 34  &  1318 \xpm 43   \\ 
Tag Efficiency     & 0.121 \xpm 0.007  &  0.368 \xpm 0.017  &  0.114 \xpm 0.006 &  0.169 \xpm 0.009 \\ \hline
\end{tabular}
\end{center}
\end{table}

\renewcommand{\secname}{Likelihood}
\section{Likelihood fit method}
\label{sec:Likelihood}

%
%
In the presence of backgrounds, the probability distribution functions 
$\cal {H}_{\pm}$ of Eq.~\ref{eq:pdf}
must be extended to include a 
term for each significant background source.
The background parameterizations are allowed to differ for each tag category.
Each event is identified as being either mixed ($-$) or unmixed ($+$)
and as belonging to
a particular tag category, $i$. Thus a distribution must be specified for 
each possibility $(+/-,i)$:
\begin{equation}
{{\cal H}_{\pm,i}} =
 f_{sig, i}{\cal {H}}_{\pm}(\dt;\Gamma,\Delta m_D,{\cal{D}}_i,\hat a_i) +
\sum_{\beta={\rm bkgd}} f_{\beta,i} {\cal{B}}_{\pm,i,\beta}(\dt;\hat b_{\pm,i,\beta})
\end{equation}
where
the background PDFs, ${\cal{B}}_{\pm,i,\beta}$, provide an empirical 
description the $\Delta t$ distribution of the background events in the sample.  
The fraction of background events for each source and tagging category is given by  
$f_{\beta,i}$, while $\hat b_{\pm,i,\beta}$ are parameters used to characterize each
source of background by tagging category for mixed and unmixed events.  The
signal fraction in each tag category is given by
\begin{equation}
f_{sig,i}=1-\sum_\beta f_{\beta,i}
\end{equation} 
The distributions are normalized so that for each $i$ and $\beta$
\begin{equation}
\int_{-\infty}^\infty d\Delta t ({\cal {B}}_{+,i,\beta} + {\cal {B}}_{-,i,\beta})=1.
\end{equation}

\subsection{Background parameterization}

Backgrounds stem from many different sources; we use control samples,
derived whenever possible from the data itself, to characterize the
background time dependence and dilution.
These control samples were described in Section~\ref{sec:Sample}.
We use an empirical description for the time dependence of the
backgrounds in the likelihood fit allowing for three time components 
for each background, each with its own dilution factor $\cal D$ and a common
resolution function $\cal R$ :

\begin{itemize}
\item   { Zero lifetime component}:
$B_{\pm ,1}=(1 \xpm {\cal{D}}_{1}^\prime )\delta(t)\otimes {\cal{R}}$
\item { Non-zero lifetime component, no mixing}:
$B_{\pm ,2}= {{\Gamma_{2}}\over 2}(1 \xpm {\cal{D}}_{2}^\prime )\
 { \rm e}^{(-\Gamma_2 | \Delta t|)} \otimes {\cal{R}}$
\item  {Non-zero lifetime component, with mixing}:
$B_{\pm ,3}={{\Gamma_3}\over2} {\rm e}^{(-\Gamma_3|\Delta t|)} 
(1\xpm {\cal{D}}_3\cos(\Delta m_3 \Delta t))\otimes {\cal{R}}$
\end{itemize}

\noindent
A likelihood fit to the background control samples is used to
determine how much of each time component is present and to fit for
the apparent lifetimes, resolution, mixing and  dilutions that
best describe the background sample. 
This approach allows for
more fit parameters than are absolutely necessary.
The goal is to determine the background 
shapes as well as possible in an empirical sense.

\subsection{Likelihood fit results}
\subsubsection{Hadronic decay modes}
We extract $\Delta m_d$ and the mistag rates by fitting the $\Delta t$ 
distributions of the selected $B$ candidate events with $\mes > 5.2$\gevcc\ 
with the likelihood function described above. 
The probability that a $B$ candidate is a signal or a 
background event is determined from a fit to the \mes\ distribution.
We describe the \mes\ shape with a single Gaussian $S(\mes)$
for the signal and the ARGUS function~\cite{ARGUS_bkgd} $B(\mes)$ for
the background. Based on this fit, the event-by-event signal
probability is determined from:  

\begin{equation}
p_{\rm sig}(\mes) = \frac{S(\mes)}{S(\mes)+B(\mes)}
\end{equation}

\noindent
The contribution of each event to the fitted signal parameters
corresponds to this signal probability.

The $\Delta t$ distributions of the combinatorial
background are described with a zero lifetime component and a 
non-oscillatory component with non-zero lifetime. 
We fit for separate resolution function
parameters for the signal and the background in order
to minimize correlations
between the background parameters and the signal parameters. 

\begin{figure}[htpb]
  \begin{center}
    \epsfig{file=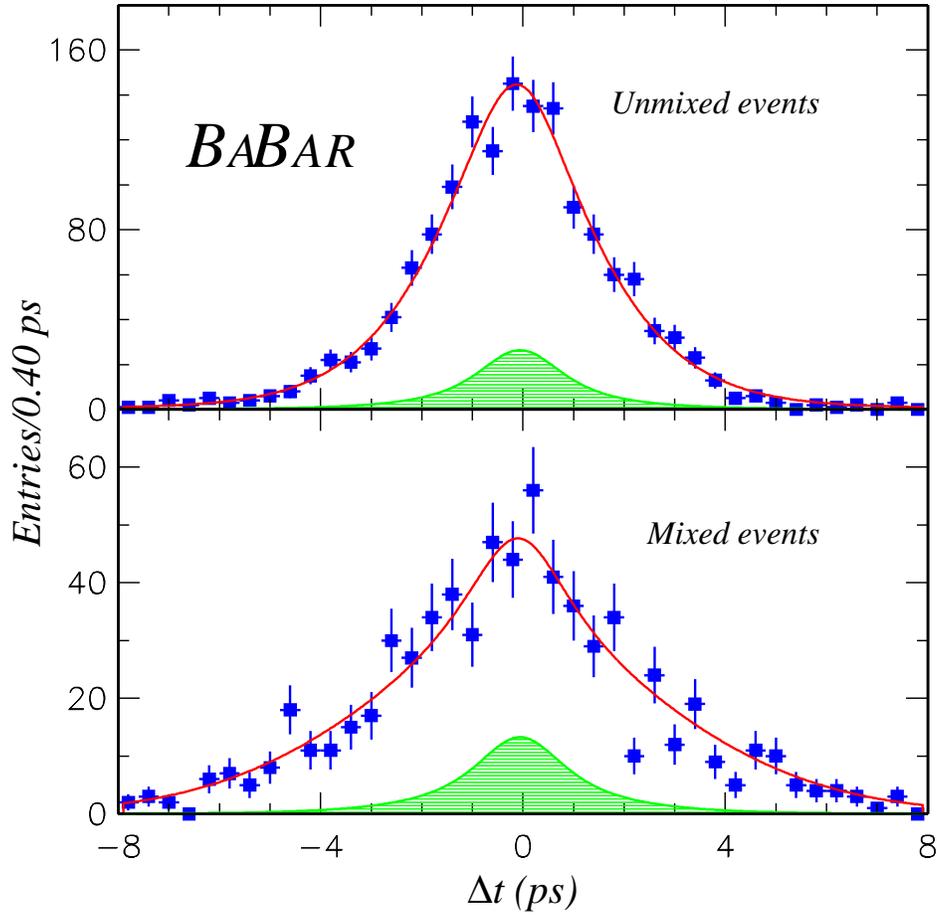,width=14.0cm}
  \end{center}
  \caption{$\Delta t$ distributions in data for the signal
    hadronic $B$ sample with $\mes > 5.27$\gevcc. The fitted $\Delta t$ 
    shapes for the selected candidates and for the fraction of background 
    candidates are overlaid. 
    The confidence level for this projection of the
    fit result is calculated from the binned $\Delta t$ distributions 
    using a Poisson-$\chi^2$ technique to be 13\% . 
   \label{fig:b0.excl-data.deltat.all}}   
\end{figure}

\begin{figure}[htpb]
  \begin{center}
    \epsfig{file=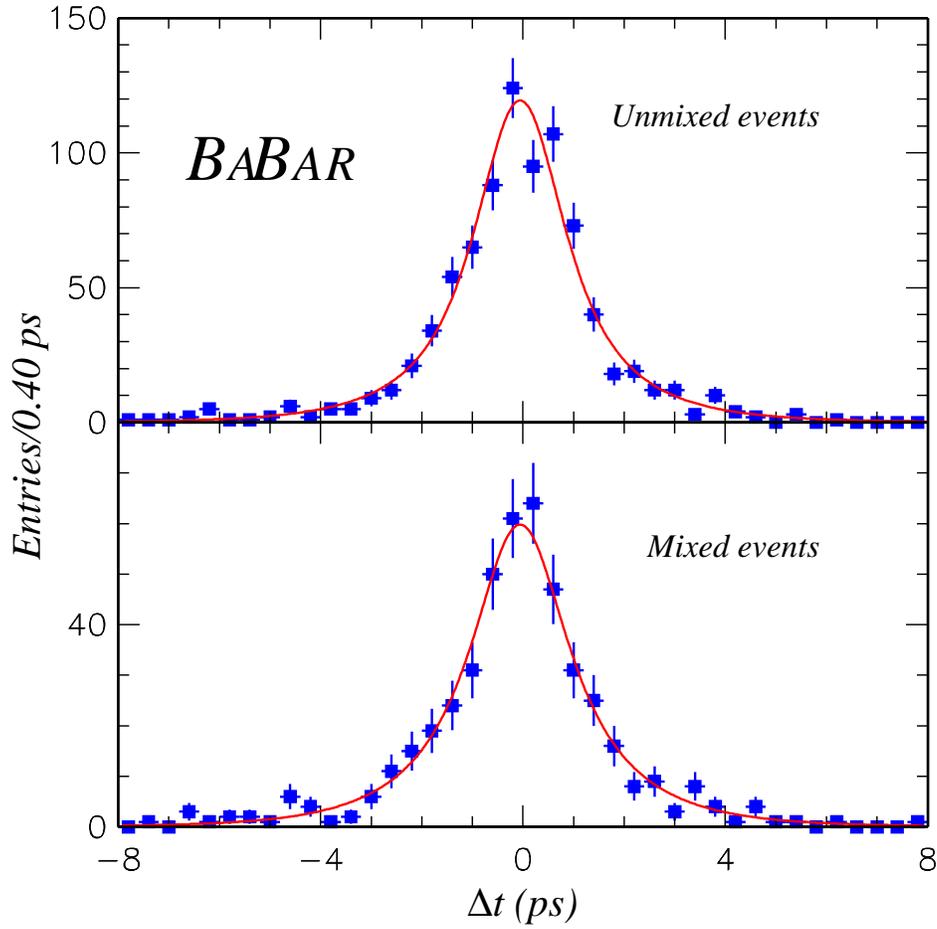,width=14.0cm}
  \end{center}
  \caption{$\Delta t$ distributions in data for the background
      candidates with $\mes < 5.27$\gevcc\ 
      separately for unmixed and mixed candidates. The fitted $\Delta t$
      shapes for those candidates are overlaid. 
      The confidence level for this projection of the
      fit result is calculated from the binned $\Delta t$ distributions 
      using a Poisson-$\chi^2$ technique to be 21\% . 
     \label{fig:b0.excl-data.deltat.all-sb}}   
\end{figure}

\begin{figure}[htpb]
  \begin{center}
    \mbox{\epsfig{file=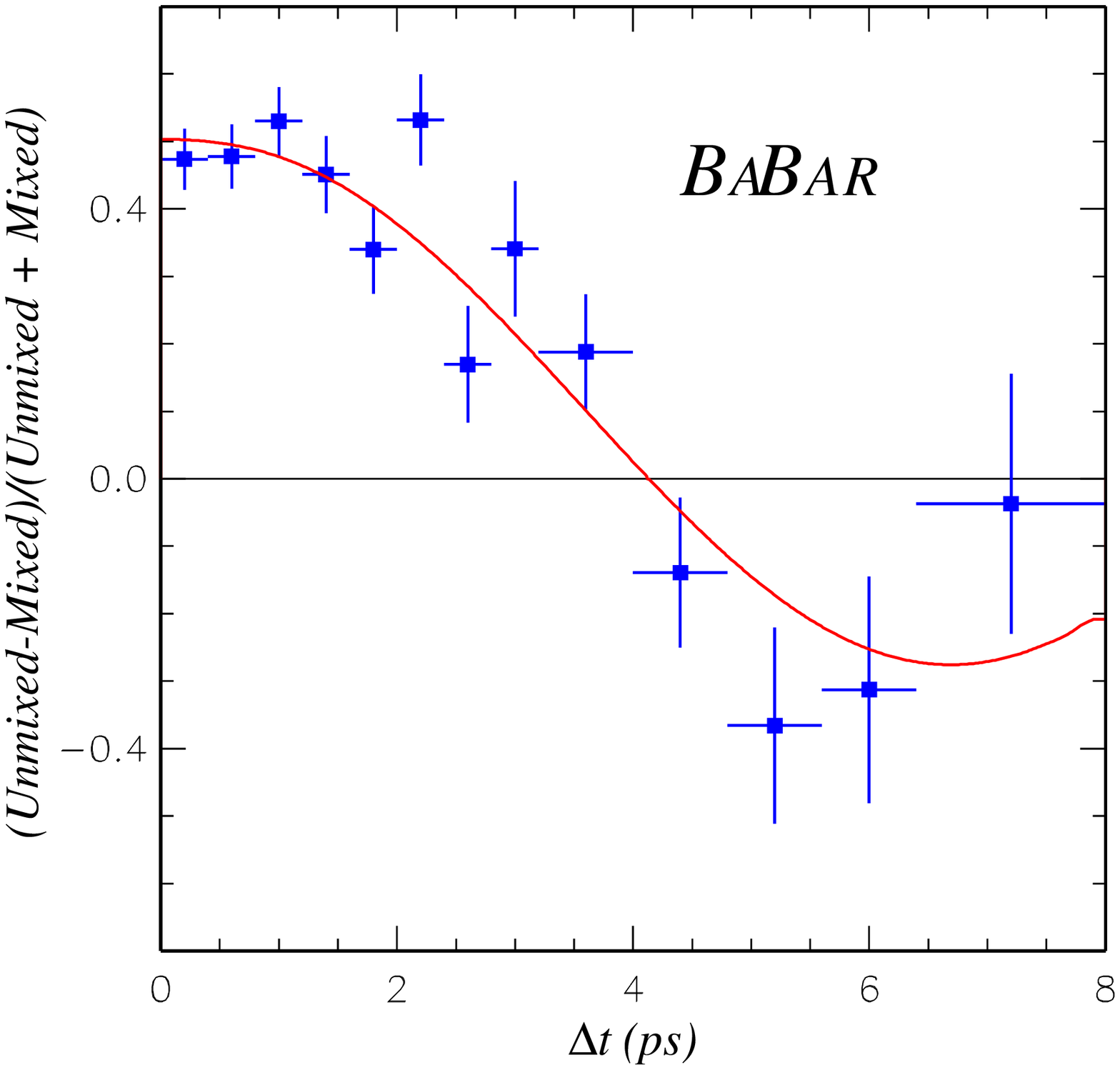,width=14.0cm}}
%
  \end{center}
  \caption{Time-dependent asymmetry $a(\Delta t)$ between unmixed and
      mixed events for hadronic $B$ candidates with $\mes > 5.27$\gevcc.
      \label{fig:b0.excl-data.cosine}}   
\end{figure}

We use the data sample described in Section~\ref{sec:Sample}, consisting
of $\approx 1900$ fully-reconstructed and tagged \Bz\ candidates.
The $\Delta t$ distributions of those candidates, overlaid with the 
likelihood fit results, are shown in Fig.~\ref{fig:b0.excl-data.deltat.all} 
for the candidates with $\mes > 5.27$\gevcc. The $\Delta t$ distributions 
for the background candidates in data from the \mes\ sideband ($\mes < 5.27$\gevcc) 
are shown in Fig.~\ref{fig:b0.excl-data.deltat.all-sb}.
In Fig.~\ref{fig:b0.excl-data.cosine} the mixing asymmetry 
of Eq.~\ref{eq:asym} is plotted;
the primordial cosine function is clearly visible. 
The results from the likelihood fit to data are summarized in
Table~\ref{tab:result-likeli}. The tagging separation
$Q=\epsilon_{tag}(1-2\mistag)^2$ is  calculated from the mistag rate
and the efficiency quoted in Table~\ref{tab:tagging-excl}.
Summing over all tagging categories, we measure a combined effective
tagging efficiency $Q\approx 28 \%$.

\begin{table}[!htb]
\begin{center}
  \caption{Results from the likelihood fit to the $\Delta t$ 
distributions of the hadronic and semileptonic $B$ decays. 
$\Delta m_d$ and 
the mistag rates include small corrections corresponding to the 
difference between the generated and reconstructed values in simulated signal events.
The summed {\it Q} over all tagging catagories is 0.285 (0.283) for hadronic (semileptonic) decay modes. } \vspace{0.3cm}
\label{tab:result-likeli}
\begin{tabular}{|l|cc|cc|}
\hline
Parameter &\multicolumn{2}{|c|}{Hadronic}
          &\multicolumn{2}{|c|}{Semileptonic} \\
          & Fit Value & $Q=\epsilon(1-2\mistag)^2$ 
          & Fit Value & $Q=\epsilon(1-2\mistag)^2$\\ 
    \hline\hline
$\Delta m_d$ [$\hbar$ ps$^{-1}$]& 0.516 \xpm\ 0.031 &
     ---                & 0.508 \xpm\ 0.020   & ---    \\
\mistag({\tt Lepton})         & 0.116 \xpm\ 0.032 & 0.062
                     & 0.084 \xpm\ 0.020 & 0.071    \\
\mistag({\tt Kaon})           & 0.196 \xpm\ 0.021 & 0.136 
                     & 0.199 \xpm\ 0.016 & 0.133    \\
\mistag({\tt NT1})            & 0.135 \xpm\ 0.035 & 0.064 
                     & 0.210 \xpm\ 0.028 & 0.066    \\
\mistag({\tt NT2})            & 0.314 \xpm\ 0.037 & 0.023 
                     & 0.361 \xpm\ 0.025 & 0.013     \\ 
   \hline
${\rm scale}_{\rm core,\ sig}$& 1.33 \xpm\ 0.13 & ---
                             &  1.32 \xpm\ 0.07 & ---\\  
$\delta_{\rm core,\ sig}$~[ps]&--0.20 \xpm\ 0.07 & ---
                             & --0.25 \xpm\ 0.04&  ---\\  
$f_{\rm outlier}$            & 0.016 \xpm\ 0.006 &  ---
                             & 0.000 \xpm\ 0.002  & ---\\  
\hline
  \end{tabular}
\end{center}
\end{table}

As a validation of these results we have carried out identical analysis
procedures on simulated Monte Carlo events generated with a detailed 
detector simulation, processed through the event reconstruction
chain in the same manner as the data. 
In the  signal simulation, the fitted values of the \BzBzb\ oscillation 
frequency, $\Delta m_d = 0.451 \xpm\ 0.011$ $\hbar$\ps$^{-1}$ (hadronic decays)
and $\Delta m_d = 0.456 \xpm\ 0.009$ $\hbar$\ps$^{-1}$ (semileptonic decays),
are consistent with the value of 0.464~$\hbar$\ps$^{-1}$ used for Monte Carlo generation. 
The fitted 
tagging dilutions and mistag rates are in good agreement with the values obtained 
from Monte Carlo truth information, confirming an unbiased measurement of those parameters. 
We apply the observed differences as a correction to the measured 
values in data, although all parameters as determined from the simulated sample are 
consistent with the generated values. 
 
\subsubsection{Semileptonic decays}
The event selection for the \btodstarlnu\ sample is summarized in 
Section~\ref{sec:Sample}.
The yield of tagged \btodstarlnu\ events and the signal purity 
are
described in Section~\ref{sec:Sample}.

The measurement of \BzBzb\ mixing and the extraction of \deltamd\
and the mistag fractions with \btodstarlnu\ decays proceeds in two
steps. First, we fit the background control samples described in
Section~\ref{sec:Sample} and determine their parameters.

\begin{figure}[htpb]
  \begin{center}
    \mbox{\epsfig{file=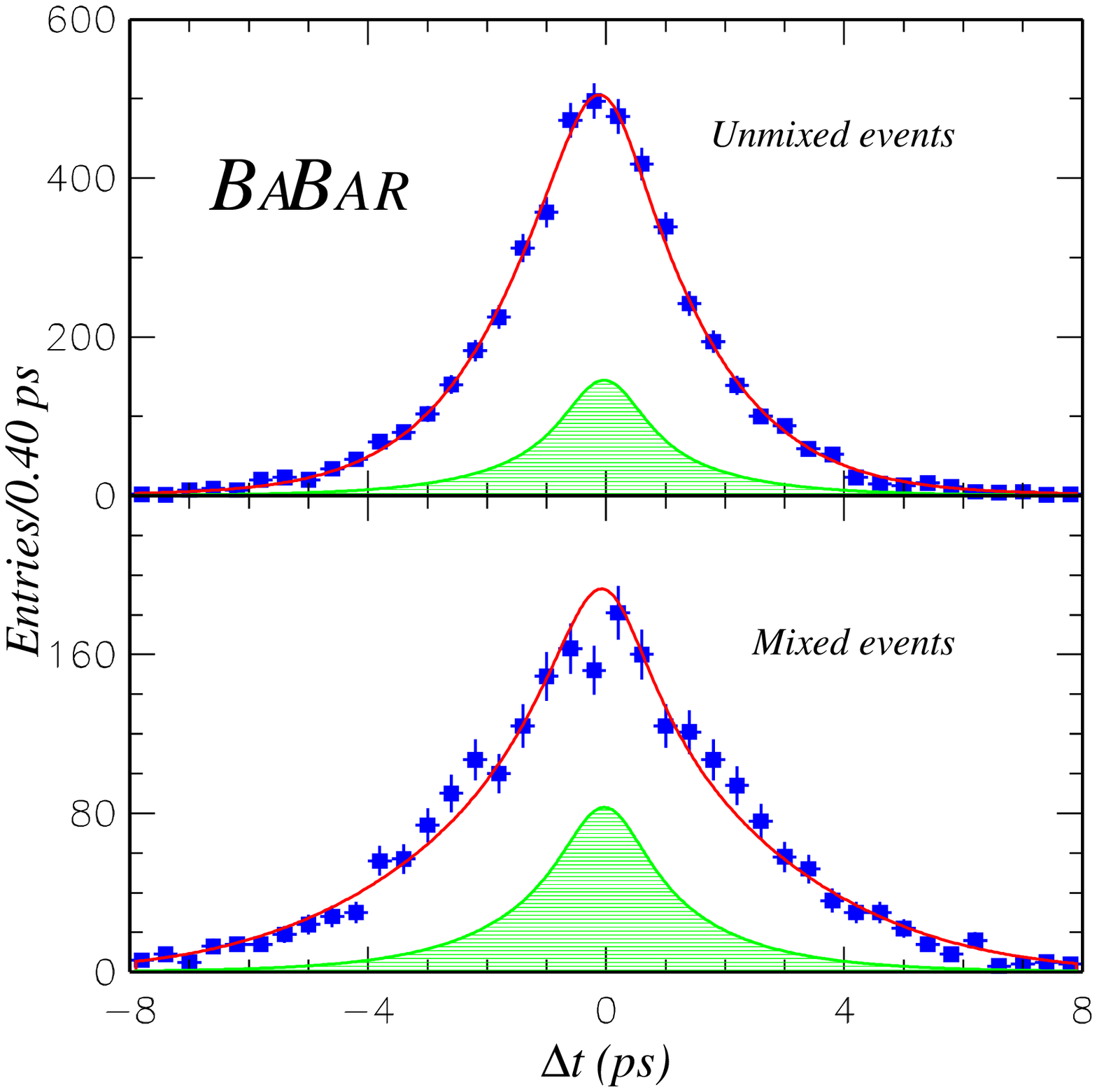,width=14.0cm}}
  \end{center}
  \caption{$\Delta t$ distributions in data for the selected \btodstarlnu\ 
      candidates for
      unmixed and mixed candidates. The fitted $\Delta t$ shapes for
      those candidates and for the fraction of background candidates are
      overlaid. 
      The confidence level for this projection of the
      fit result is calculated from the binned $\Delta t$ distributions 
      using a Poisson-$\chi^2$ technique to be 28\%. 
   \label{fig:b0mix.dslnu-data.deltat.cats}}  
\end{figure}

Second, we fit the signal
events, fixing the background fractions to the values summarized in
Table~\ref{tab:D*lnu_bkgd} and the $\Delta t$
resolutions and dilution parameters  
to the values obtained from fits to the control samples. 
The results of the fit and the calculated tagging performance,
$Q=\epsilon_{tag}(1-2\mistag)^2$, are summarized in 
Table~\ref{tab:result-likeli}.
The $\Delta t$ distribution of the \btodstarlnu\  candidates, overlaid with the 
likelihood fit result, is shown in Fig.~\ref{fig:b0mix.dslnu-data.deltat.cats}


\begin{figure}[htpb]
  \begin{center}
    \mbox{\epsfig{file=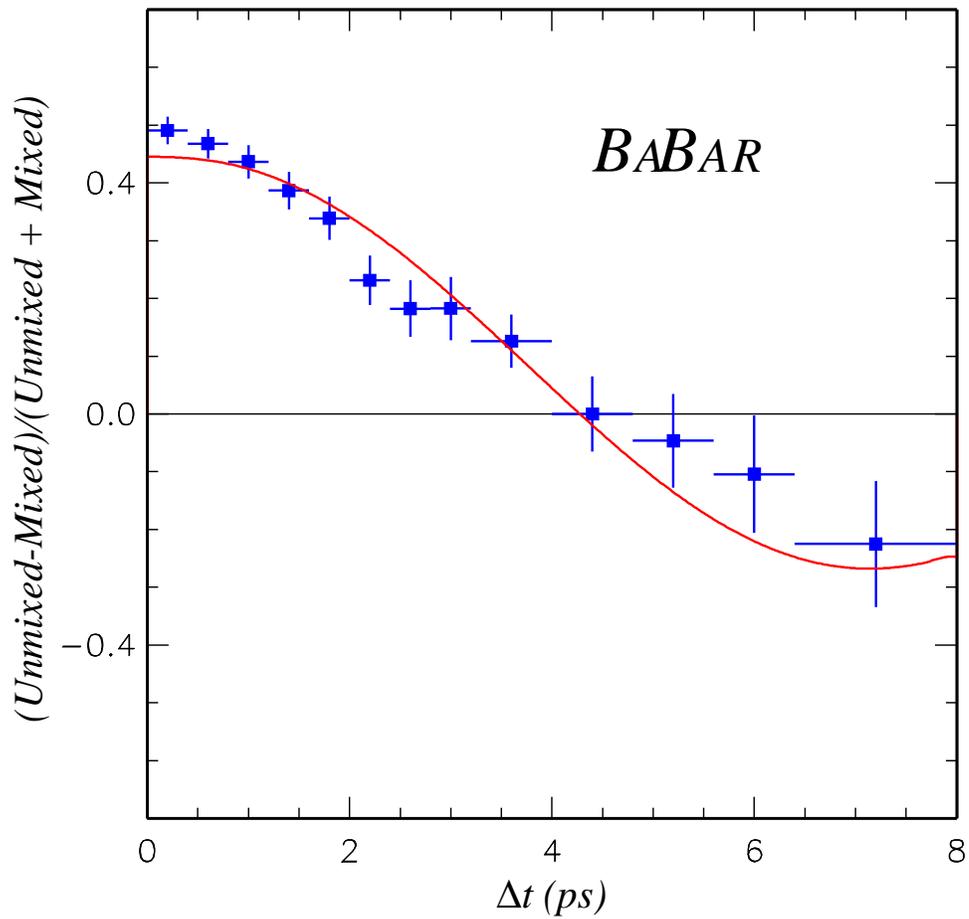,width=14.0cm}}
  \end{center}
  \caption{Time-dependent asymmetry between unmixed and mixed events
    for \btodstarlnu\ candidates.\label{fig:b0.dslnu-data-mc.cosine}}    
\end{figure}

\subsection{Systematic error estimation} 
The systematic errors for the hadronic and semileptonic $B$ samples
are summarized in Tables~\ref{tab:syst-had} and \ref{tab:syst-semil} and
can be grouped into three categories:

\bigskip
{\noindent\bf \boldmath $\Delta t$ reconstruction}
\medskip

We determine the level of any potential systematic bias in \deltamd\ and the mistag rates 
due to the $\Delta t$ resolution function by independently varying the scale factor
and the mean of the wide Gaussian and the fraction of events in the wide Gaussian. 
The variations  correspond to a change in the RMS of the total resolution 
function by one standard deviation as measured in data. 
The systematic uncertainty due to $\Delta t$ outliers is estimated from 
the variation of the fitted parameters with the fraction of outliers and 
the RMS of their $\Delta t$ distribution in a sample of toy Monte Carlo events. 
An error in the boost of the $\Upsilon(4S)$ system or in the $z$ scale of the 
detector can bias the \deltamd\ measurement because those parameters are used to
reconstruct a decay length difference $\Delta z$ and to convert it to the decay time 
difference $\Delta t$. In the likelihood fit, we fix the $B^0$ lifetime 
to the PDG value~\cite{PDG} and its uncertainty leads to a systematic error.

\bigskip
{\noindent\bf Background parameters}
\medskip

The signal probability assigned to each candidate in the hadronic $B$ 
sample has a statistical uncertainty, and these statistical uncertainties lead to
systematic uncertainties in \deltamd\ and the mistag rates. We estimate these
uncertainties by varying the width and height of the fitted peak in \mes\
by one standard deviation.

In the semileptonic sample we vary the average 
background fractions by the statistical uncertainties derived from the control 
samples. The uncertainty in the fraction of $D^*\ell\nu$ candidates with a fake lepton
includes the uncertainty in the lepton identification rates.
To estimate the sensitivity of \deltamd\ and the mistag rates to the $\Delta t$ 
description of the combinatorial background, we repeat the likelihood fit with an 
additional oscillatory term in the PDF. In the fit to the signal sample we vary the 
background dilutions obtained from the control samples by one standard deviation.
We study the sensitivity to the resolution function of the combinatorial
backgrounds by allowing an additional scale factor to account for possible
tails in the $\Delta t$ distribution.

\bigskip
{\noindent\bf Check with simulated events}
\medskip
 
Candidate selection criteria can cause systematic biases
in the measurement of \deltamd\ and the mistag rates. These biases are
estimated with fully simulated events and are found to be consistent with zero
within their statistical uncertainty. Nevertheless,
we correct for the actual differences between the generated and reconstructed values and 
include the statistical uncertainties in the measured parameters from the simulated events as 
a contribution to the systematic uncertainties in the result.


\begin{table}[htbp]
\begin{center} 
 \caption{Systematic uncertainties for \deltamd\ and the mistag
    rates measured with hadronic $B$ decays for the likelihood fit.\label{tab:syst-had}}  \vspace{0.3cm}
  \begin{tabular}{|l|ccccc|}\hline
    Source  & $\Delta m_d$ &{\tt Lepton}&{\tt Kaon}&{\tt NT1}& {\tt NT2} \\ 
    & [$\hbar$ ps$^{-1}$]  & & & & \\
    \hline\hline
    $\Delta t$ Resolution  & 0.011 & 0.004 & 0.004 & 0.004 & 0.004 \\
        \hline
    Background $\Delta t$  & 0.002 & 0.002 & 0.002 & 0.002 & 0.002 \\
    Background Resolution  & 0.002 & 0.002 & 0.002 & 0.002 & 0.002 \\
    Background Fractions   & 0.004 & 0.004 & 0.002 & 0.006 & 0.004 \\ 
        \hline
    $B^0$ lifetime         & 0.005 & 0.001 & 0.001 & 0.001 & 0.001 \\
    $z$ scale              & 0.005 & --- & --- & --- & --- \\     
    $z$ boost              & 0.003 & --- & --- & --- & --- \\     
        \hline  
    Monte Carlo Correction          & +0.013 & --0.001 & 0.000 & --0.010 & --0.015 \\
                           &\xpm\ 0.011&\xpm\ 0.011&\xpm\ 0.008&\xpm\ 0.015&
                                                            \xpm\ 0.014\\  
    \hline\hline
    Total Systematic Error & 0.018 & 0.013 & 0.010 & 0.017 & 0.015 \\
    Statistical Error      & 0.031 & 0.032 & 0.021 & 0.035 & 0.037 \\
    \hline\hline
    Total Error            & 0.036 & 0.035 & 0.023 & 0.039 & 0.040 \\
        \hline  
   \end{tabular}
\end{center}
\end{table}

\begin{table}[htbp]
\begin{center} 
 \caption{Systematic uncertainties in \deltamd\ and in the mistag
    rates measured with semileptonic $B$ decays for the likelihood fit.\label{tab:syst-semil}}  \vspace{0.3cm}
  \begin{tabular}{|l|ccccc|}\hline
    Source  & \deltamd\ & {\tt Lepton} &{\tt Kaon} &{\tt NT1}&{\tt NT2} \\ 
    & [$\hbar$ ps$^{-1}$] & & & & \\
    \hline\hline
    $\Delta t$ Resolution  & 0.012 & 0.005 & 0.009 & 0.012 & 0.005 \\ \hline
    Background $\Delta t$  & 0.002 & 0.002 & 0.002 & 0.002 & 0.002 \\
    Background Resolution  & 0.002 & 0.002 & 0.002 & 0.002 & 0.002 \\
    Background Dilutions   & 0.006 & 0.008 & 0.013 & 0.026 & 0.031 \\
    Background Fractions   & 0.006 & 0.009 & 0.011 & 0.017 & 0.032 \\
    $B^+$ Backgrounds      & 0.010 & 0.009 & 0.010 & 0.004 & 0.003 \\
        \hline
    $B^0$ lifetime         & 0.006 & 0.001 & 0.001 & 0.001 & 0.001 \\
    $z$ scale              & 0.005 & --- & --- & --- & --- \\     
    $z$ boost              & 0.003 & --- & --- & --- & --- \\     
        \hline  
    Monte Carlo Correction          & +0.008 & --0.010 & --0.001 & --0.002 & --0.006   \\
                  & \xpm\ 0.009 & \xpm\ 0.008 & \xpm\ 0.006 & \xpm\ 0.011 & \xpm\ 0.011   \\  
    \hline\hline
    Total Systematic Error & 0.022 & 0.018 & 0.023 & 0.035 & 0.046 \\ 
    Statistical Error      & 0.020 & 0.020 & 0.016 & 0.028 & 0.025 \\
    \hline\hline
    Total Error            & 0.030 & 0.027 & 0.031 & 0.045 & 0.052 \\ 
\hline
  \end{tabular}
\end{center}
\end{table}

\renewcommand{\secname}{Onebin}
\section{Time-integrated method}
\label{sec:Onebin}  

\subsection{Description of the method}
As previously described in Section~\ref{sec:Introduction}, a
time-integrated method to measure the mistag fractions in data
provides a simple and robust check of the likelihood fit method
presented in Section~\ref{sec:Likelihood}. The statistical precision
of this method is enhanced by restricting the sample to
events  in a single optimized $\Delta t$ interval.
Taking into account the \babar\ vertex resolution, the
optimum interval is found to be  $|\Delta t| < 2.5$\ps.  
This is so because the number
of mixed events in this time interval is dominated by the mistag
rate rather than by  \BzBzb\  mixing.  Events with $|\Delta t|> 2.5$\ps\
have on average equal numbers of mixed and unmixed events due to
\BzBzb\ oscillations, and therefore
contribute nothing to the determination of the mistag rate.
We refer to this time-integrated
method using a single optimized $\Delta t$ interval as the ``single-bin''
method.

The single-bin analysis uses the reconstructed  hadronic and semileptonic 
\Bz\ sample described in Section~\ref{sec:Sample}.  The number of
tagged events in each category were summarized in Table~\ref{tab:tagging-excl} 
and
Table~\ref{tab:D*lnu_bkgd}.  The background fractions were re-evaluated
for the sample of tagged signal events with $|\Delta t| < 2.5$\ps.  

To correct for the presence of backgrounds, we must add to
Equation~\ref{eq:TagMix:Integrated} a term to account for the
contribution of
each background source to the fraction of mixed events in the sample:
\begin{equation}
\label{eq:onebin_bkgd}
\chi_{obs} = f_s (\chi_d  + (1 - 2 \chi_d )\, \mistag) + 
\sum_{\beta} f_{\beta} \chi_{\beta}, 
\end{equation}
where $f_s, f_{\beta}$ are the fraction of signal and each background
source, respectively, $\chi_{\beta}$ is the fraction of
mixed events in each background source, and $\chi_{obs}$ is the
observed fraction of mixed events.  
We restrict the sample to events with $|\Delta t| < 2.5$\ps; then 
$\chi_d$ must be modified to represent the integrated mixing
probability for $|\Delta t| < 2.5$ ps.  Using the world-average value
for \deltamd, $0.472 \pm 0.017~\hbar $ ps$^{-1}~$\cite{PDG}, and
taking into account the 
detector resolution function ${\cal{R}}(\Delta t)$, we find
\begin{equation}
\chi_d^{\prime} = \frac{1}{2} \lbrack 1 - \frac
{\int_{-2.5 ps}^{+2.5 ps}\{e^{-|\dt|/\tau}cos(\deltamd \dt) \otimes {\cal{R}}(\dt) \}d(\dt)}
{\int_{-2.5 ps}^{+2.5 ps}\{e^{-|\dt|/\tau} \otimes {\cal{R}}(\dt)\}d(\dt)} \rbrack
                 = 0.079.
\end{equation}

\noindent
Solving for Eq.~\ref{eq:onebin_bkgd} for $\mistag$, and using the calculated
value for $\chi_d^{\prime}$, we obtain
\begin{equation}
\label{eq:onebin_w}
\mistag = {{\chi_{obs} - f_s \chi_d^{\prime}  - \sum_{\beta} f_{\beta} \chi_{\beta}}
\over {f_s (1 - 2 \chi_d^{\prime} )} } .
\end{equation}

\subsection{Results}
\subsubsection{Hadronic sample}
The selection of hadronic $B$ mesons was described in 
Section~\ref{subsec:Sample_hadronic}.
We use all tagged events with  $|\Delta t| < 2.5$\ps,
and determine the background fraction
in the signal sample from a fit to the \mes\ distribution
as described in Section~\ref{sec:Likelihood}.  The signal
region is defined as events with $\mes > 5.27 $\gevcc.
The fraction of
mixed events in the background is determined by tag category
using the sideband
control sample, $\mes < 5.27$\gevcc.


We use Eq.~\ref{eq:onebin_w} to solve for the mistag rate for each tag
category, obtaining the results shown in Table~\ref{tab:onebin_w}.

\begin{table}[htbp]
\begin{center}
  \caption{Mistag rate $\mistag$ and tagging separation $Q$ 
as measured by the single-bin method in the hadronic
and semileptonic $B$  event samples. 
The mistag rates include small corrections corresponding to the 
difference between the generated and reconstructed values in simulated signal events.
\label{tab:onebin_w}}   \vspace{0.3cm}
  \begin{tabular}{|l|cc | cc|}\hline
    Tagging Category & \multicolumn{2}{|c|}{Hadronic} & 
    \multicolumn{2}{|c|}{Semileptonic} \\
 & {Mistag Rate \mistag } & $Q=\epsilon_{tag}(1-2\mistag)^2$ 
 & {Mistag Rate \mistag } & $Q=\epsilon_{tag}(1-2\mistag)^2$ \\
    \hline\hline
{\tt Lepton}    & 0.120\xpm\ 0.032   &0.061 
                & 0.095\xpm\ 0.018   &0.079          \\
{\tt Kaon}      & 0.207\xpm\ 0.021   &0.126
                & 0.177\xpm\ 0.014   &0.154          \\
{\tt NT1}       & 0.127\xpm\ 0.035   &0.067 
                & 0.200\xpm\ 0.024   &0.041        \\
{\tt NT2}       & 0.342\xpm\ 0.036   &0.016
                & 0.346\xpm\ 0.024   &0.016        \\
    \hline
    All & ---   & 0.270  & --- & 0.290    \\
\hline
  \end{tabular}
\end{center}
\end{table}

\subsubsection{Semileptonic sample}
The selection of \btodstarlnu\ events was described in 
Section~\ref{subsec:Sample_semileptonic}.
We use tagged events with $|\Delta t| < 2.5$\ps\ and re-evaluate
the backgrounds for events in this time interval, with the control samples
described above.  The backgrounds are evaluated for each tag category and
for each $D^0$ decay mode.  The mistag fractions are calculated
individually by tag category and decay mode using Eq.~\ref{eq:onebin_w},
and the results for the different decay modes are combined, using 
the statistical errors to weight the individual results.
All systematic errors are conservatively taken to be correlated
between the different decay modes.
The determination of systematic errors will be discussed
in the next section.
The results are
summarized in Table~\ref{tab:onebin_w}.


\subsection{Systematic errors}

The various sources of systematic error in the single-bin method
that have been investigated 
for the hadronic and semileptonic $B$ samples are 
summarized in  Tables~\ref{tab:syst_onebin_had} and 
\ref{tab:syst_onebin_semil}
and discussed in detail below.

\begin{table}[htbp]
\begin{center}
\caption {Sources of systematic error for the mistag measurement
on the hadronic sample in the single-bin method.
See text for details.\label{tab:syst_onebin_had}} \vspace{0.3cm}
  \begin{tabular}{|c|c|c|c|c|c|}
\hline
Type       & Variation &{\tt Lepton}&{\tt Kaon}& {\tt NT1} & {\tt NT2}  \\
\hline\hline
Background &$\pm1\sigma$ & 0.004    &0.004     &0.008      &0.006  \\
$\chi_d$ &$0.174\pm0.009$& 0.005    &0.004     &0.005      &0.004   \\
Resolution& see text     & 0.002    &0.002     &0.002      &0.002  \\
wrong tag resolution& see text   & 0.005 &0.009  &0.007 &0.013  \\
Monte Carlo correction&              & +0.009&+0.004 &-0.026&-0.007 \\
Monte Carlo statistics&              & 0.012 &0.008  &0.017 &0.015 \\
\hline
Total           &           & 0.015 &0.014  &0.021 &0.021 \\
\hline
\end{tabular}
\end{center}
\end{table}

\begin{table}[htbp]
\begin{center}
\caption {Sources of systematic error for the mistag measurement
from the semileptonic sample in the single-bin method.
See text for details.\label{tab:syst_onebin_semil}} \vspace{0.3cm}
  \begin{tabular}{|c|c|c|c|c|c|}
\hline
Type    & Variation & {\tt Lepton} & {\tt Kaon} & {\tt NT1} & {\tt NT2}  \\
\hline\hline
Background  & $\pm 1\sigma$ & 0.008 &0.010  &0.023 &0.020  \\
$B^-\rightarrow D^{*-}X \ell^+ \nu$& see text      & 0.009 &0.009  &0.009 &0.008   \\   
$\chi_d$    & $0.174\pm0.009$& 0.012 &0.009  &0.009 &0.004   \\
Resolution          & see text & 0.002 &0.002  &0.002 &0.002 \\  
wrong tag resolution& see text & 0.005 &0.009  &0.009 &0.013  \\
Monte Carlo correction&                 &0.001  &-0.001 &-0.003&-0.014 \\
Monte Carlo statistics&                 & 0.010 & 0.006 &0.013 & 0.012  \\
\hline
Total &                     & 0.021 &0.020  &0.031 &0.028 \\
\hline
\end{tabular}
\end{center}
\end{table}

The systematic error due to background in the data samples is taken
by varying both the background fractions and the fraction of mixed events
associated to each background source,
$\chi_{\beta}$.  These quantities are varied by one
standard deviation of the values measured in the background control
samples described in Section~\ref{sec:Sample}.  
For the semileptonic sample,
this is the dominant source of systematic error, primarily due to the
limited statistics of the background control samples. 

The systematics uncertainties introduced by background from
the decay $B^+\rightarrow D^{*-}X \ell^+ \nu$ are obtained by varying 
the fraction described in Section~\ref{subsec:Sample_semileptonic}
as well as the mistag fraction of the charged $B$ meson measured on data.

The assumed $\Delta t$ resolution function is a double-Gaussian
where the initial parameters are taken from simulation.  The two widths
are  multiplied by two scale factors which are determined in a fit 
to the data along with the fraction of events in the narrow Gaussian. 
The double-Gaussian RMS is consistent with that of the resolution
function considered in Section~\ref{sec:Likelihood}.  These three 
parameters are varied in a conservative way 
within the errors of the fit to the data
to determine the systematic uncertainty on the mistag fraction, which
is quite small.

Finally, we consider the possibility that wrong tags have worse
$\Delta t$ resolution than correct tags.  This effect has
been studied in Monte Carlo simulation, where we observe a slightly larger RMS 
in the $\Delta t$ distribution for events with wrong-sign tags.
We assign a systematic error  on the mistag rates 
by taking the default resolution function 
and changing the  parameters of the resolution function for wrong tags
by a corresponding amount.

\renewcommand{\secname}{Compare}
\section{Comparison of likelihood and single-bin methods}
\label{sec:Compare}  

Combining the results obtained for the hadronic and semileptonic 
$B$ samples for the likelihood fit method described in  
Section~\ref{sec:Likelihood} and for the single-bin method described
in Section~\ref{sec:Onebin}, and
taking into account the correlated systematic errors, we obtain
the preliminary mistag rate results summarized in Table~\ref{tab:mistag-all}.
The Monte Carlo corrections to the likelihood fit results summarized
in Tables~\ref{tab:syst-had} and \ref{tab:syst-semil} have been
applied to the likelihood fit results of Table~\ref{tab:result-likeli}.

\begin{table}[htbp]
\begin{center}
  \caption{Combined mistag results for the hadronic and
semileptonic $B$ samples, for the likelihood and single-bin methods.} \vspace{0.3cm}
\label{tab:mistag-all}
  \begin{tabular}{|l|c|c|}\hline
   Tag Category &\multicolumn{2}{|c|}{Mistag, \mistag} \\
                & Likelihood     &  Single-Bin   \\
\hline\hline

{\tt Lepton} & 0.096 \xpm\ 0.017 \xpm\ 0.013 &  0.102\xpm\ 0.016 \xpm\ 0.015 \\
{\tt Kaon}   & 0.197 \xpm\ 0.013 \xpm\ 0.011 &  0.187\xpm\ 0.012 \xpm\ 0.015 \\
{\tt NT1}    & 0.167 \xpm\ 0.022 \xpm\ 0.020 &  0.176\xpm\ 0.020 \xpm\ 0.024 \\
{\tt NT2}    & 0.331 \xpm\ 0.021 \xpm\ 0.021 &  0.345\xpm\ 0.020 \xpm\ 0.023 \\

    \hline
  \end{tabular}
\end{center}
\end{table}

The single-bin fit results use a sub-sample of the events used in the
likelihood fit, so the two sets of results
are correlated. The
systematic errors differ, due to the different sensitivities of the
two methods.  Overall, the two methods agree well.
The effective flavor tagging efficiency
of the algorithm described in Section~\ref{sec:Tagging} is found
to be $Q \approx 0.28$.


\renewcommand{\secname}{Summary}
\section{Summary}
\label{sec:Summary}

In  8.9\invfb\ of \epem\ annihilation data collected near the \FourS\ resonance, 
we have obtained a preliminary measurement of the time-dependent \BzBzb\ oscillation frequency using a sample of \Bz\ mesons 
reconstructed in hadronic decay channels and in the semileptonic decay mode  $\Bz\to
D^{*-}\ell^+\nu$.

From the hadronic \Bz\ sample we measure the \BzBzb\ oscillation frequency:
\begin{eqnarray*}
  \Delta m_d    =  0.516\ \xpm\ 0.031\ ({\rm stat.})\  \xpm\ 0.018 \ ({\rm syst.})\ \hbar \ {\rm ps}^{-1}
\end{eqnarray*}

\noindent
From the $D^{*-}\ell^+\nu$ sample we measure the \BzBzb\ oscillation frequency:

\begin{eqnarray*}
  \Delta m_d    =   0.508\ \xpm\ 0.020\ ({\rm stat.})\ {}\xpm\ 0.022\ ({\rm syst.})\ \hbar \ {\rm ps}^{-1}
\end{eqnarray*}

\noindent
Combining the \deltamd\ results from the hadronic and semileptonic
$B$ samples, we obtain the preliminary result : 
\begin{eqnarray*}
\Delta m_d = 0.512 \xpm\  0.017 \ ({\rm stat.})\  \xpm\ 0.022\ ({\rm syst.})\  \hbar \ \mbox{ps}^{-1}.
\end{eqnarray*}

\noindent
In combining the two results, we have taken all systematic error contribution to be 
fully correlated with the exception of the contribution due to Monte Carlo simulation 
statistics.

\section{Acknowledgments}
\label{sec:Acknowledgments}


We are grateful for the contributions of our \pep2\ colleagues in
achieving the excellent luminosity and machine conditions
that have made this work possible.
We acknowledge support from the
Natural Sciences and Engineering Research Council (Canada),
Institute of High Energy Physics (China),
Commissariat \`a l'Energie Atomique and
Institut National de Physique Nucl\'eaire et de Physique des Particules
(France),
Bundesministerium f\"ur Bildung und Forschung
(Germany),
Istituto Nazionale di Fisica Nucleare (Italy),
The Research Council of Norway,
Ministry of Science and Technology of the Russian Federation,
Particle Physics and Astronomy Research Council (United Kingdom), the
Department of Energy (US),
and the National Science Foundation (US). In addition, individual support 
has been received from the Swiss 
National Foundation, the A. P. Sloan Foundation, the Research Corporation,
and the Alexander von Humboldt Foundation.
The visiting groups wish to thank 
SLAC for the support and kind hospitality
extended to them.


\end{document}